\documentclass{article}

\newcommand*\der{\mathop{}\!\mathrm{d}}
\usepackage{graphicx}
\usepackage{algorithm}
\usepackage{algpseudocode}
\usepackage{multirow}
\usepackage{float}
\usepackage{longtable}
\usepackage{epstopdf}
\usepackage{xcolor}
\usepackage{multicol}
\usepackage[colorlinks,bookmarksopen,bookmarksnumbered,citecolor=blue,urlcolor=blue]{hyperref}
\usepackage[misc]{ifsym}
\usepackage{pdflscape}
\usepackage{mathtools, nccmath}
\usepackage[utf8]{inputenc}  
\usepackage{natbib}    
\usepackage[english]{babel}

\usepackage[letterpaper,top=2cm,bottom=2cm,left=3cm,right=3cm,marginparwidth=1.75cm]{geometry}

\usepackage[defaultmathsizes,italic]{mathastext}
\usepackage{amsmath, amsfonts}
\usepackage{booktabs}

\makeatletter
\newcommand{\evalat}[2]{\mathpalette\eval@at{{#1}{#2}}}
\newcommand{\eval@at}[2]{\eval@@at#1#2}
\newcommand{\eval@@at}[3]{%
  #2\,
  {%
   \sbox0{$#1\left|#2\right|$}%
   \vrule height \dimexpr\dp0+0.1ex depth \dimexpr\dp0+0.5ex\relax
  }_{\,#3}
}

\date{}

\title{U-DESPE: a Bayesian Utility-based methodology for dosing regimen optimization in early-phase oncology trials based on Dose-Exposure, Safety, Pharmacodynamics, Efficacy}
\author{A. Andrillon (1,2), S. Micallef (3), M. Ursino (4), P. Mozgunov (5), M-K Riviere (1)\\
\small{1) Department of Statistical Methodology, SARYGA, France}\\
\small{2) INSERM U1153 Team ECSTRRA, Université Paris Cité, Paris, France}\\
\small{3) Debiopharm International SA, Lausanne, Switzerland} \\
\small{4) HeKA, INSERM, Inria, Université Paris Cité, France}\\
\small{5) MRC Biostatistics Unit, University of Cambridge, UK}}

\begin{document}
\maketitle

\begin{abstract}
With the development of novel therapies such as molecularly targeted agents and immunotherapy, the maximum tolerated dose paradigm that ``more is better'' does not necessarily hold anymore. In this context, doses and schedules of novel therapies may be inadequately characterized and oncology drug dose-finding approaches should be revised. This is increasingly recognized by health authorities, notably through the Optimus project.
We developed a Bayesian dose-finding design, called U-DESPE, which allows to either determine the optimal dosing regimen at the end of the dose-escalation phase, or use of dedicated cohorts for randomizing patients to candidate optimal dosing regimens after that safe dosing regimens have been found. U-DESPE design relies on a dose-exposure model built from pharmacokinetic data using non-linear mixed-effect modeling approaches. Then three models are built to assess the relationships between exposure and the probability of selected relevant endpoints on safety, efficacy, and pharmacodynamics. These models are then combined to predict the different endpoints for every candidate dosing regimens. Finally, a utility function is proposed to quantify the trade-off between these endpoints and to determine the optimal dosing regimen. We applied the proposed method on a clinical trial case study and performed an extensive simulation study to evaluate the operating characteristics of the method.  \end{abstract}

\section{Introduction}\label{intro}
Historically, phase I cancer dose-finding clinical trial designs were developed for cytotoxic drugs and aimed at identifying the maximal tolerated dose (MTD) of a new compound defined as the highest dose with acceptable toxicity. This paradigm has been challenged with the arrival of new classes of anticancer agents, such as molecularly-targeted therapies (MTAs) and immunotherapies. Indeed, their modes of action target the immune system or tumor-specific pathways and differ from those of cytotoxic chemotherapies. \cite{Hoering2011, PostelVinay2011}.  In this sense, their action may not be linked to cell destruction, and the assumption used for cytotoxic agents that the more toxic, the more efficacious the drug does not appear relevant anymore. When increasing the dose, a plateau in efficacy may occur before reaching non tolerable dose. Therefore, MTD may not be the optimal dose and the pharmacology of this class of agents is important to consider when selecting the dose for later development. 
 In 2021, the U.S. Food and Drug Administration (FDA) Oncology Center of Excellence initiated the Project Optimus to reform the dose optimization in early phase oncology\cite{project_optimus}. This project, led by a wide range of experts (including oncologists, pharmacologists, toxicologists, and statisticians) highlighted that doses and schedules of novel therapies may be inadequately characterized using the MTD paradigm and that oncology drug dose-finding approaches should be revised. In August 2024, the FDA issued guidance for industry \cite{guidance_optimus}, that provides recommendations to identify the ‘optimal dose’, that is a dose regimen that optimizes the risk–benefit trade-off of the drug or provides desired therapeutic effect whilst minimizing toxicity. The dose-optimization should be based on relevant non-clinical and clinical data (such as pharmacokinetics, pharmacodynamics, safety, tolerability, and activity), as well as the dose- and exposure-response relationships. Moreover, one of the key objectives of Project Optimus is to favor early and comprehensive multiple dosing regimen exploration during the clinical development of oncology drugs to support the proposed recommended dosing regimen(s). Sponsors and researchers are encouraged to compare rigorously different doses at the end of phase I trials, or in phase II trials evaluating the efficacy of at least two doses identified in phase I, through randomized trials. \\ 

The idea of using multiple endpoints relevant for dose selection (usually toxicity and either efficacy or pharmacokinetic) is not new, but comprehensiveness and balance of the endpoints has not been much supported by rigorous approaches.  In the literature, a large number of designs were developed for dose selection based on both toxicity and efficacy endpoints. For instance, Thall and Russell\cite{Thall1998} developed the first model-based design for toxicity and efficacy endpoints. Braun \cite{braun2002} proposed a bivariate continual reassessment method for both toxicity and disease progression with an association coefficient constant across doses. Cunanan and Koopmeiners \cite{cunanan2014} compared copula models for the joint probability of toxicity and efficacy with independent models. Asakawa et al. \cite{asakawa2014} proposed to use Bayesian model averaging for bivariate continual reassessment to accommodate the uncertainty in the skeleton specification. Ivanova\cite{ivanova2003} proposed an algorithm for trinary outcome with the objective to maximize the probability of response without toxicity. Thall and Cook introduced the EffTox design~\cite{thall2004}, a Bayesian adaptive dose-finding design based on trade-offs between the probabilities of efficacy and toxicity. Colin et al.\cite{Colin2017} proposed a Bayesian approach based on a two parameters (toxicity-activity) utility function. Two-stage phase I/II designs, such as the approach by Hoering et al.\cite{Hoering2011} and the Dose-Ranging Approach to Optimizing Dose (DROID)~\cite{Guo2023}, were also developed. These two-stage designs first identify the MTD through conventional dose limiting toxicity (DLT) based dose-escalation, and then randomize patients among multiple doses to identify the optimal dose based on risk–benefit assessment.  \\

Some authors have also developed designs that incorporate pharmacokinetic (PK) data, typically using a measure of exposure, to model toxicity and/or efficacy endpoints \cite{ursino2017,micallef2022,pantoja2022,takeda2018,gerard2022, su2022,yang2024,yuan2024}. Lu et al.\cite{Lu2024} proposed a Bayesian PK integrated dose-schedule finding design to identify the optimal dose regimen by integrating PK, toxicity, and efficacy data. Yuan et al. recently \cite{yuanOpt2024} provided a review of different design strategies for dose-optimization trials, including phase I/II designs. However, there is a need for a unified approach to support dosing regimen recommendations on the basis of any relevant perspectives, i.e. balancing safety, pharmacodynamic activity and efficacy, while taking into account the exposure and its variability. Our approach aims to create a supporting tool to quantify the benefit of each dosing regimen, balancing the desirable and non desirable effects of every candidate regimens. Usually, desirable effects are related to the pharmacodynamic activity and efficacy, while the non desirable effects are linked to the toxicity. In the context of early development, where little is known on the compound's effect on these processes and only limited number of patients are usually treated, relationships are difficult to quantify and the approach cannot be solely empirical. The mechanism of action of the compound is also major to be assessed. Therefore, the pharmacology and sources of variability have also to be handled in the process of dose optimization using modeling tools.\\

In this paper, we develop a Bayesian Utility-based approach for dosing regimen optimization relying on Dose-Exposure- Safety, Pharmacodynamics, Efficacy relationships modeling for oncology early phase clinical trials, named U-DESPE. The proposed design aims to address the following objectives: (1) at the end of a dose-escalation part, determine candidate(s) optimal dosing regimen(s) to be further explored, (2) at any time after exploring safe dose levels, randomize the next cohorts of patients to optimal regimen candidates and (3) support the selection of the dosing regimen with quantitative and rigorous tools. This approach is based on the assessment of the compound's pharmacology. A non-linear mixed effect model describes the PK relationship from which relevant measures of exposure can be derived and empirical models are constructed to quantify the exposure-safety/pharmacodynamics/efficacy relationships. These models are combined to predict the corresponding probabilities of events for every candidate dosing regimen. A utility function is proposed to quantify the trade-off between these endpoints and is maximized to determine the optimal dose level. This methodology is motivated by an ongoing clinical trial, and we performed an extensive simulation study to evaluate the operating characteristics of the proposed method.\\

\section{Motivating example}\label{sec:motivating}

The present work is motivated by a phase I dose-escalation and dose-expansion study which primary objective was to determine the recommended phase 2 dose (RP2D) of Debio 0123 in combination with Carboplatin (NCT03968653,\cite{Debio0123_101}). This study  evaluates the safety, tolerability, pharmacokinetics, and anti-tumor activity of an oral, highly selective brain penetrant inhibitor of the tyrosine kinase WEE1 (Debio 0123) in adult patients with advanced solid tumors. In this multicenter, open-label, phase I dose-escalation study, Debio 0123 is administered in adults with selected recurrent/progressive solid tumors. Two dosing regimens were tested in two different dose-escalation cohorts:  D1-D3 given over a 21-day cycle (corresponding to arm A) or D1-D3 and D8-D10 over a 21-day treatment cycle (corresponding to arm B). The main objective of the study was to identify the MTD of the two dosing regimens and select the recommended phase II dosing regimen (RP2D). Accrual of the dose-escalation is completed ; 38 and 17 patients were treated respectively in arm A and arm B based on Bayesian dose-escalation approaches. The study results showed that both dosing regimens of Debio 0123 in combination with Carboplatin were well tolerated, with a manageable safety profile and the MTD was established as 520 mg with both dosing regimens. Increasing exposure was observed in the plasma of patients treated with increasing Debio 0123 doses. Pharmacodynamic (PDy) analysis demonstrated target engagement, as characterized by reductions in p-CDC2 levels in skin biopsies following treatment, at $\ge$ 200 mg doses, further supporting the established RP2D. In exploratory analyses of this unselected cohort of patients, a trend was seen between Debio 0123 plasma exposure and both p-CDC2 levels and the antitumor activity of Debio 0123. RP2D selection was based on the incidence of DLTs and cumulative safety data. Other secondary endpoints, such as antitumor activity, PK, and PDy, and exposure relationships were also taken into account for the selection of the RP2D. The U-DESPE handling PK, PDy, safety and efficacy was used at the end of the dose-escalation phase to determine a range of doses to be further explored as potential optimal doses and to support the RP2D selection.\\

Concurrently with the debio0123.101, the debio0123.102 study (NCT05109975\cite{Debio0123_102}) was conducted as an open-label, multicenter phase 1 study assessing the safety, preliminary antitumor activity, PK, and PDy of Debio 0123 monotherapy. The primary objectives of the dose-escalation part of this trial were to establish the MTD and the RP2D. Twenty-seven patients with relapsed/refractory advanced solid tumors were treated with Debio 0123 at doses ranging from 30–350 mg according to a Bayesian Logistic Regression Model escalation \cite{Neuenschwander2008}. In this study, Debio 0123 was given orally, once daily, and demonstrated a manageable toxicity profile. Consistently with the debio0123.101 findings, pharmacodynamic (PDy) data showed target engagement through reductions in p-CDC2 levels in skin biopsies. These two studies collected data that could potentially reinforce the knowledge on each other; particularly, information related to pharmacokinetics and pharmacodynamics can be shared under certain verifiable assumptions - such as minimal impact of Carboplatin on Debio 0123's pharmacokinetics and no Carboplatin exposure at the time of target engagement (p-CDC2) measurement.  Borrowing information via a common analysis of relevant data can be beneficial and can support a better selection of the candidate optimal dosings.\\

 \begin{figure}[h]
\centerline{\includegraphics[scale=0.65]{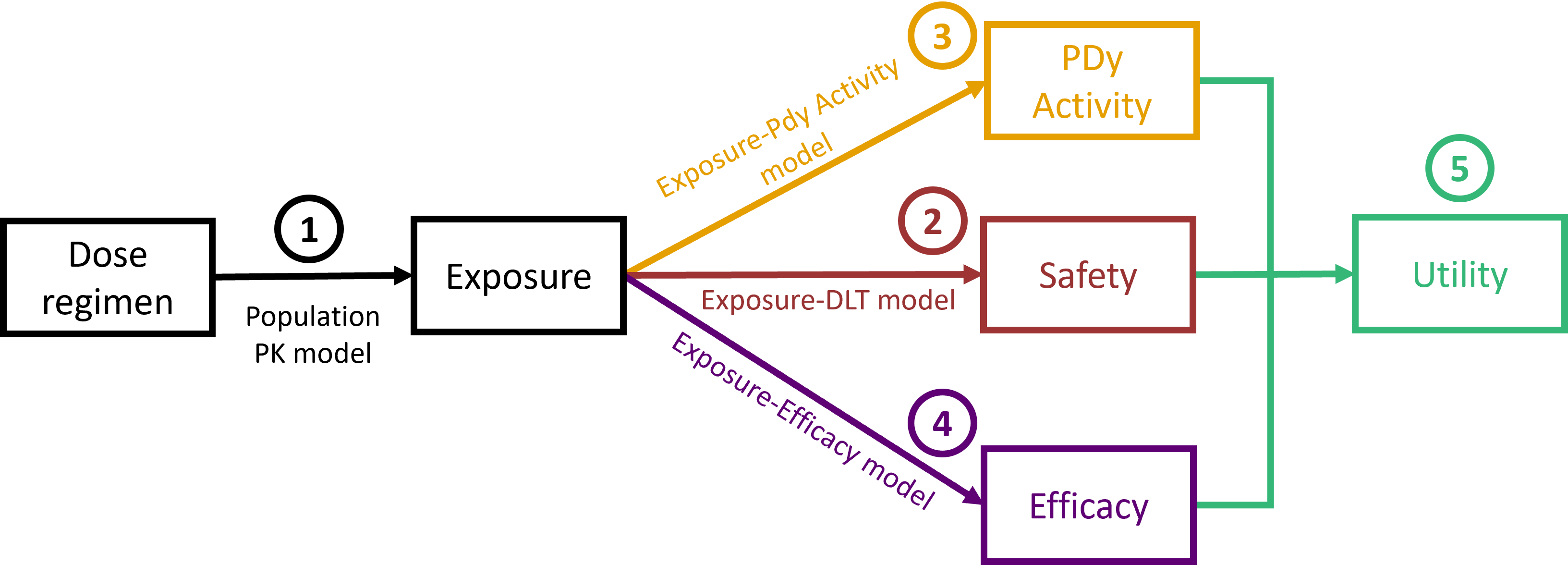}}
\caption{General modeling approach composed of (1) PK modeling to describe exposure as a function of dosing regimens, (2)-(4) models to explain safety, pharmacodynamic activity, and efficacy according to exposure, and (5) a utility function to integrate the different endpoints and provide a global recommendation.}\label{fig:Global_modeling}
\end{figure}

Discussions with the extended clinical team and exploratory analyses suggested that safety, PDy, and efficacy may be better characterized by exposure, which can vary between patients receiving the same dosing regimen, rather than by the dose itself. Therefore, our modeling approach was divided into several parts as described in Figure~\ref{fig:Global_modeling}.
First, we used a population PK model to describe the relationship between the dosing regimens and the concentration and derive exposure measures. Then, we built different exposure-response models for the endpoints of interest: the safety (usually evaluated through DLTs), the PDy activity (p-CDC2 relative decrease) and the efficacy (anti-tumor activity assessed via tumor shrinkage). If relevant, it is possible to consider different measures of exposure for each endpoint. Finally, the utility function is computed from these endpoints in order to quantify the benefit-risk balance of every dosing regimen.\\

\section{Methods}\label{sec:method}
We consider a dose-finding clinical trial where a set of $J$ dosing regimens of the compound are investigated, each dosing regimen being defined by a dose amount, an administration schedule, and a duration of treatment\cite{regimen_NCI}. Let $(D_j)_{j=1,...,J}$ denote the different dosing regimens, and $D_i$ represents the dosing regimen administered to patient $i$. Similar notations will be used for the different outcomes considered.

\subsection{Population pharmacokinetic (PopPK) model}\label{sec:popPK-model}

A nonlinear mixed-effects model is used to describe the continuous drug concentration $C(t,D)$ at time $t$ with dosing regimen $D$ as follows: 
\begin{equation} \label{eq:popPK-model}
C(t,D) = f(\theta_i,t,D) + h(\theta_i,t,D,\xi) \times \varepsilon
\end{equation}

\noindent where $f$ represents the structural PK model which is usually compartmental, and in such case $f$ is the solution of  a set of differential equations with random parameters. Let $\theta_i$ represents the i$^{th}$ patient's specific parameter vector, where $\theta_i=\mu e^{(\eta_i)}$, with $\mu$ denoting the fixed effects and $\eta_i$ denoting the random effects defined as $\eta_i \sim \mathcal{N}(0,\Omega)$ with $\Omega$ denoting the variance-covariance matrix. The error model is represented by $h$ which depends on additional parameter $\xi$, and $\varepsilon$ is a standard Gaussian variable. The most usual error models are the constant model where $h(\theta_i,t,D,\xi=a)=a$, the proportional model where $h(\theta_i,t,D,\xi=b)=bf(\theta_i,t,D)$, and combinations of the constant and proportional models.\\

To be consistent with the motivating example, we used a one-compartment with first order absorption and linear elimination PK model, parameterized by the absorption rate constant ($k_a$), clearance ($CL$), and volume of distribution ($V$), that is, using the same notation as in Equation (\ref{eq:popPK-model}),
\begin{equation}
\label{eq_conc}
    f \left( \theta_i=(k_{a,i}, CL_i, V_i), t, D=[(d)_{\times L}] \right) = \sum_{\substack{\ell = 1\\ t_{\ell} \leq t}}^L \frac{d}{V_{i}} \frac{k_{a,i}}{k_{a,i} - k_{i}} \left(e^{-k_{i} \left(t - t_{\ell} \right)} - e^{-k_{a} \left(t - t_{\ell} \right)}\right),
\end{equation}
where $k_i=CL_i/V_i$ for patient $i$, and $\ell=1, ..., L$ refers to the $L$ administrations of dose level $d$ at time $t_{\ell}$ from dosing regimen $D$. 
Lognormal random effects were applied to $k_a$ and $CL$ to account for inter-individual variability (IIV), while a proportional error model was used to describe residual variability, that is, $k_{a,i} \sim LN\left(\log(k_{a,pop}), \sigma_{k_a}^2\right)$, ${CL}_i \sim LN\left(\log(CL_{pop}), \sigma_{CL}^2\right)$, $V_i = V_{pop}$ and $b=1$ (cf. Equation (\ref{eq:popPK-model})).
The parameters $k_{a,pop}$, $CL_{pop}$, and $V_{pop}$ represent the mean values at the population level for absorption rate, clearance, and volume of distribution, respectively, while $\sigma_{k_a}^2$ and $\sigma_{CL}^2$ denote the IIV for the corresponding parameters.

Based on PK sample collected in patients, the population and individual parameters can be estimated. Several individual exposure measures can then be derived, such as the area under the curve over 24 hours (AUC24), the cumulative AUC over the first cycle ($AUC_{cum}$), the maximal  concentration ($C_{max}$), and the trough concentration reached by the drug immediately before the next dose is administered ($C_{trough}$). We denote by $Z_i$ the individual exposure metric of patient $i$ relevant to explain the endpoint of interest, noting that different measures can be relevant for different endpoints. The relevant exposure measure will then be linked to each considered endpoint via an exposure-effect model in the next subsections.

\subsection{Bayesian exposure-safety model}\label{sec:DLT-model}

We assume that toxicity is evaluated through the dose limiting toxicity  defined by the trial investigator and is monotonically increasing with the exposure. Let $W_i$ denote the binary toxicity outcome of patient $i$, with $W_i=1$ (respectively $W_i=0$) indicating a DLT (respectively no DLT), and $\evalat{p_i}{Z_i} =P(W_i=1|Z_i)$ denote the DLT probability of patient $i$ with individual exposure metric $Z_i$. We model toxicity using a Bayesian Logistic Regression Model (BLRM)\cite{Neuenschwander2008,gerard2022}, but considering a relevant exposure metric instead of dose levels, as follows:
\begin{equation} \label{eq:toxicity-model}
logit\left( \evalat{p_i}{Z_i} \right)=\phi_1+e^{\phi_2}\log\left(\frac{Z_i}{Z_{ref}}\right)
\end{equation}
where $\phi_1$ and $\phi_2$ are unknown parameters, and $Z_{ref}$ is a reference exposure. We consider a Bivariate Normal prior distribution $\phi= (\phi_1, \phi_2) \sim BVN(\mu_\phi,\Sigma_\phi)$. 

\subsection{Bayesian exposure-PDy activity model}\label{sec:activity-model}

Let denote $\evalat{R_i}{Z_i}$ the continuous PDy activity outcome of patient $i$ with individual exposure metric $Z_i$. The exposure-PDy activity relationship is defined by a general Bayesian model $g$ that fits the best to the study data:
\begin{equation} \label{eq:PD-model}
\evalat{R_i}{Z_i} = g(\beta, Z_i) + \zeta_i
\end{equation}
with $\beta$ the parameter vector to be estimated, and $\zeta_i$ an error term with $\zeta_i \sim \mathcal{N}(0,\sigma_r^2)$.\\

In our work, we selected a specific model for $g(.)$ that is appropriate given our motivating example and the type of data to be collected. However, any model that is suitable and consistent with the characteristics of the trial data under consideration could be employed. Indeed, in our motivating example, the PDy activity is evaluated using  a marker (referenced as p-CDC2) collected in skin tissue and measured by immunochemistry techniques. The drug is expected to target the protein kinase WEE1 that phosphorylates (and thus inactivates) cell division cycle 2 (CDC2). Therefore, the p-CDC2 is considered as a PDy marker, and the endpoint for PDy activity is the H-Score percent change from baseline, calculated as the difference between the p-CDC2 H-score in skin biopsy at pre-treatment (baseline) and on-treatment. To model the relative reduction from baseline of the PDy marker as a function of exposure, we propose a log-linear model that aligns with the range of values anticipated for $Z_i$.
\begin{equation}\label{eq:pCDC2-model}
\evalat{R_i}{Z_i} = \beta_1 + \beta_2 \log\left(\frac{Z_i}{Z_{ref}}\right) + \zeta_i
\end{equation}

We define a Bivariate Normal prior distribution on parameter $\beta= (\beta_1, \beta_2) \sim BVN(\mu_\beta,\Sigma_\beta)$ and a Gamma prior on the precision $1/\sigma_r^2 \sim \Gamma(a_0,b_0)$. \\

We may be directly interested in the exposure-pharmacodynamic modeling in order to maximize the relative reduction from baseline of the PDy. In the setting of the motivating example, a PDy signal was considered relevant when the relative reduction from baseline exceeds a certain level. Therefore, to measure the target engagement, we estimate $q_i$ the probability for patient $i$ of having a pharmacodynamic marker reduction exceeding a pre-specified threshold $c$:
\begin{equation} \label{eq:target-engagement-model}
\evalat{q_i}{Z_i} = \Pr\left( \evalat{R_i}{Z_i}\ge c \right)
\end{equation}

\subsection{Bayesian exposure-efficacy model}\label{sec:efficacy-model}

Let $\evalat{S_i}{Z_i}$ denote the continuous efficacy outcome of patient $i$ with individual exposure metric $Z_i$, described by the general Bayesian model $g$ that fits the best to the study data:
\begin{equation} \label{eq:efficacy-model}
\evalat{S_i}{Z_i} = g(\gamma, Z_i) + \zeta_i
\end{equation}
with $\gamma$ the model vector parameters to be estimated and $\zeta_i$ an error term with $\zeta_i \sim \mathcal{N}(0,\sigma_s^2)$.\\

In the motivating example, the efficacy is assessed according to the anti-tumor activity. The endpoint considered is defined as the best change (maximum reduction) in the sum of the longest diameters of the target lesions (i.e. tumor shrinkage) from the start of study treatment until disease progression or recurrence, the start of a new systemic therapy, patient discontinuation from study or analysis cut-off, whichever occurs first.  Best percentage change from baseline is then calculated as the best change from baseline divided by the tumor size at baseline. Tumor shrinkage related endpoint is used in our motivating example but any relevant efficacy endpoint can be considered and modeled for a different study. We choose to model the exposure-tumor shrinkage relationship using piecewise cubic monotone splines:

\begin{equation} \label{eq:shrinkage-model}
\evalat{S_i}{Z_i} = \gamma_0 + \sum^{L}_{l=1}\gamma_l I_{l}\left(\frac{Z_i}{Z_{ref}}\right) + \zeta_i
\end{equation}
with the spline coefficients $\gamma_0 \in \mathbb{R}$, $\forall l \in [1,L], \gamma_l \in \mathbb{R}^{+}$, $I_{l}$ the $l$'th member of a family of cubic I-splines functions and $L$ the number of degrees of freedom \cite{ramsay1988}. We set one knot for a small sample size and then increased the number of knots with the number of patients. A Gamma prior is defined on the precision $1/\sigma_s^2 \sim \Gamma(a_0,b_0)$ and on parameter $\gamma_l \sim \Gamma(\tilde{a_0},\tilde{b_0})$ to ensure an increasing exposure-efficacy relationship, and a Normal prior is used for $\gamma_0$.

\subsection{Posterior distribution of each endpoint according to the dosing regimen}\label{sec:exposure-to-dose}

The exposure-response models could enable an exposure recommendation based on some target criteria. However, exposure recommendation is not convenient for clinical practice and the objective is rather to provide a dosing regimen recommendation. Therefore, the exposure-response models are combined with the dose-exposure model to predict the corresponding response for each candidate dosing regimen\cite{micallef2022}, and by this way recommend a dosing regimen handling the variability in exposure.\\

Let $\der P(Y|D)$ denote the probability distribution of random variable $Y$ conditionally on variable $D$. According to the law of total probability, we can derive the posterior probability distribution of each endpoint $Y$ given the dosing regimen $D$:

\begin{equation} \label{eq:exposure-to-dose}
\der P(Y | D)  = \int_{Z} \der P(Y | Z, D)  \der P(Z | D)
\end{equation}
with $Y \in \{\text{W, R, S}\}$ the endpoint for either safety, PDy, or efficacy, $Z$ the exposure, and $D$ the dosing regimen. \\

The distribution $\der P(Z | D)$ is given by the PK model, while the distribution $\der P(Y | Z, D)$ is the posterior distribution of the endpoint in the exposure-endpoint model that actually does not depend on the dosing regimen. Hence, to sample in the posterior distribution of the endpoint given the dosing regimen, we first sample the exposure in the PK model, and then sample the endpoint in the posterior distribution of the exposure-endpoint model according to the previously sampled exposure (this requires, as an intermediate step, to sample in the posterior distribution of the exposure-endpoint model parameters). Thus we obtain, at each dosing regimen $j$, the estimated safety $\left( \evalat{p}{D_j} = P(W=1|D_j) \right)$, the estimated PDy $\left( \evalat{r}{D_j} \right)$ and PDy exceeding a threshold $c$ $\left( \evalat{q}{D_j} = P\left( \evalat{r}{D_j} \geq c \right) \right)$, and the estimated anti-tumor activity $\left( \evalat{s}{D_j} \right)$. For instance, for safety, Equation (\ref{eq:exposure-to-dose}) translates into:
\begin{equation} \label{eq:exposure-to-dose-safety}
\evalat{p}{D}  = \int_{\Theta} \int_{\Phi} \evalat{p}{Z(\theta, D)}(\phi) L(\phi|W,Z(\theta,D)) g(\phi) \der\phi \times Z(\theta, D) \der\theta
\end{equation}
\noindent where $Z(\theta, D)$ is the derived exposure metric from the PK model in function of parameters $\theta \in \Theta$ and dosing regimen $D$, $\evalat{p}{Z(\theta, D)}(\phi)$ is the probability of DLT from Equation (\ref{eq:toxicity-model}) in function of exposure and parameters $\phi \in \Phi$, $L(\phi|W,Z(\theta,D))$ is the likelihood of $\phi$ and $g(\phi)$ is the prior distribution of $\phi$.\\

\begin{figure}[h!]
\centerline{\includegraphics[scale=0.55]{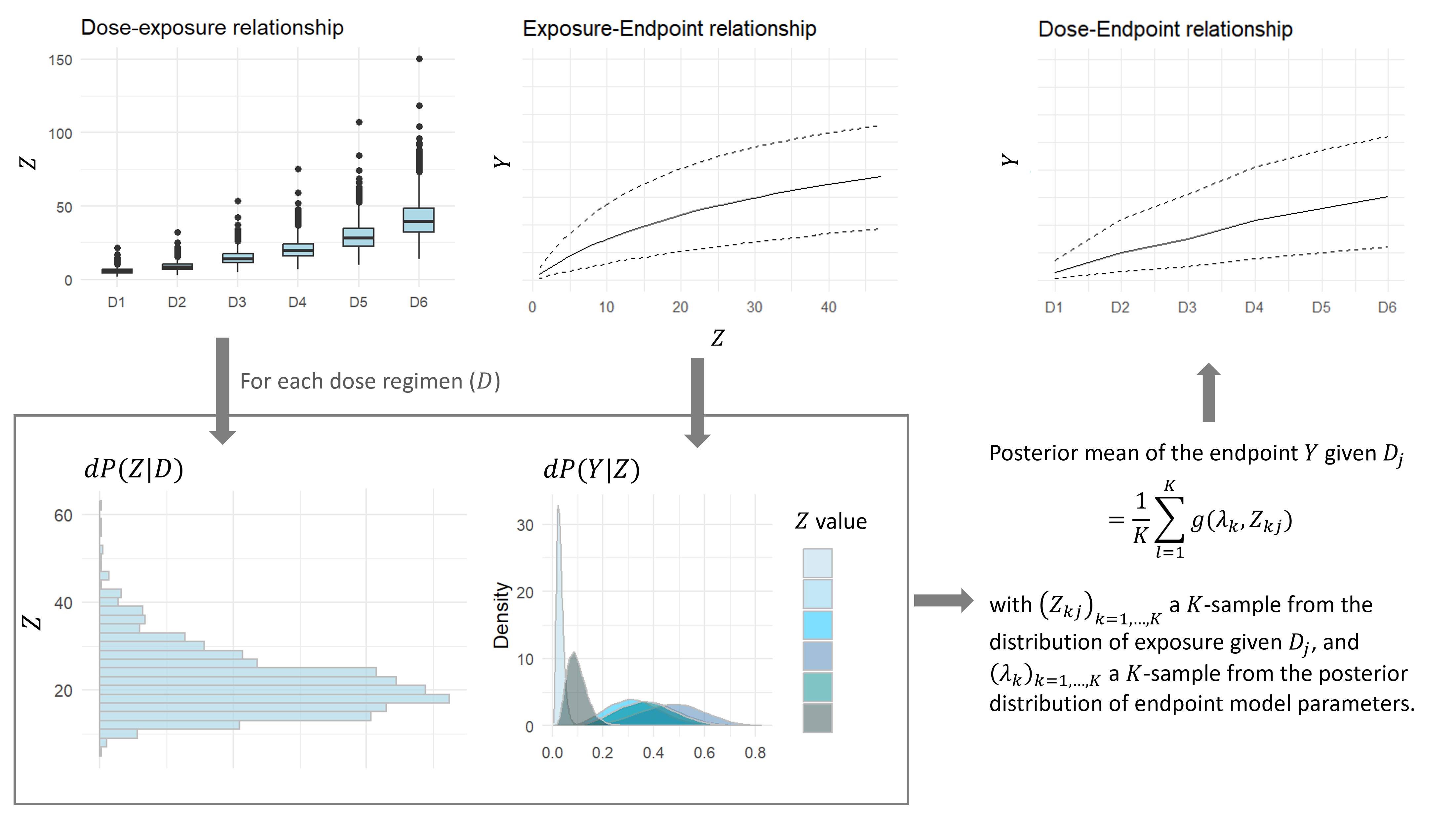}}
\caption{Exposure-to-dose shifting process for a given endpoint (either DLT, efficacy or PDy). Left panel: (top) simulations of individual exposure from the population PK model at each dosing regimen evaluated. (bottom) Distribution of the predicted exposures in the population for a given dosing regimen. Middle panel: (top) Estimated posterior probability of the endpoint, defined as a function of exposure. Solid line is for the endpoint mean estimate and dashed lines represent the 95\% prediction interval. (bottom) Histograms of the estimated distribution of endpoint rates for different exposures. Right panel: posterior probability of endpoint for each candidate dosing regimen.\label{fig:Process-dose-exp}}
\end{figure}

To illustrate practically how the dose-endpoint relationship is obtained, an example for a generic endpoint (potentially representing either DLT, efficacy, or PDy activity) is presented in Figure \ref{fig:Process-dose-exp}. On the left side of the figure, we predicted, from the PK model, the distribution of exposure for each of the dosing regimen considered. In our case, we used the area under the concentration curve over 24 hours at steady-state (AUC24). However, depending on the relevance, for each specific endpoint, any PK metric could also be utilized. In the middle part of the Figure~\ref{fig:Process-dose-exp}, the distribution of the endpoint conditional to exposure is estimated by fitting an exposure-endpoint model. On the right, we combine these distributions to obtain, given the dosing regimen, the distribution for the endpoint; this distribution handles the uncertainties and the variability inherent to PK. Let us note that in our illustration, the candidate dosing regimens differ only by the dose, but our approach is applicable for a set of dosing regimen candidates combining any dose and schedule.

\subsection{Utility function}\label{sec:utility}

The dosing regimen-endpoint relationships can be integrated into a utility function (also referred to as a gain function in this manuscript) to translate their benefit-risk and support a unique recommendation for the optimal dosing regimen.

In line with the setting of the motivating example, we proposed a gain function $G$, built from the elicitation of the clinicians' knowledge, defined as follows:  
\begin{equation} \label{eq:gain}
G\left( \evalat{p}{D_j},\evalat{q}{D_j},\evalat{s}{D_j}; \alpha_1,\alpha_2,\alpha_3,\delta_{\text{min}},\delta_{\text{max}} \right) = \left\{
    \begin{array}{ll}
    -\infty & \text{if } \evalat{p}{D_j} \ge \delta_{\text{max}} \\
    \alpha_1 \evalat{s}{D_j} + \alpha_2 \evalat{q}{D_j} + \alpha_3 \left( \evalat{p}{D_j}-\delta_{\text{min}} \right)  & \text{if } \delta_{\text{min}} \le \evalat{p}{D_j} \le \delta_{\text{max}}  \\
    \alpha_1 \evalat{s}{D_j} + \alpha_2 \evalat{q}{D_j}  & \text{otherwise} 
    \end{array}
\right.
\end{equation}
with $\evalat{p}{D_j}$ the probability of experiencing a DLT at dosing regimen $D_j$, $\evalat{q}{D_j}$ the probability that the relative decrease of PDy marker exceeds a given threshold $c$, $\evalat{s}{D_j}$ the tumor shrinkage, and $\alpha=(\alpha_1,\alpha_2,\alpha_3)$, the gain functions parameters to be calibrated. A specific weight can be assigned to each endpoint according to the value of the $\alpha$ parameters. In particular, $\alpha_1>0$  and $\alpha_2>0$ respectively favor tumor shrinkage and PDy activity, while $\alpha_3<0$ penalizes toxicity. The intervals of interest are determined based on the probability of DLT: doses levels associated with a probability of DLT exceeding a threshold $\delta_{\text{max}}$ are considered as unsafe, and $-\infty$ is assigned to the gain function, while those associated with a probability of DLT below $\delta_{\text{max}}$ represent the safe and potentially effective doses. Below the threshold $\delta_{\text{min}}$, no penalty is applied to the gain function based on the probability of DLT and the gain is a linear combination of the tumor shrinkage and the target engagement. Within the range $[\delta_{\text{min}};\delta_{\text{max}}]$ we penalize the gain according to an increasing probability of DLT. In practice, parameters are calibrated based on clinicians' preferences, who can order endpoints according to their importance for the dose determination. Statisticians can then propose scenarios to elicitate clinicians' preferences, allowing the gain parameter values to be derived iteratively.

For ease of notation, gain at dosing regimen $D_j$ will be denoted $G_j$ with \\ $G_j = G\left( \evalat{p}{D_j},\evalat{q}{D_j},\evalat{s}{D_j}; \alpha_1,\alpha_2,\alpha_3,\delta_{\text{min}},\delta_{\text{max}} \right)$.

\subsection{Dose recommendation}\label{sec:dose-reco}

Dosing regimen recommendation is performed 
through the gain function defined previously. The proposed method is intended to be applied either at the end of a dose-escalation phase to recommend an optimal dosing regimen for subsequent phases, or at any time after identifying safe dosing regimens to randomize the next cohorts of patients to potential optimal dosing regimens. 

Since some dosing regimens can have very close gain values, we would like to select lower (\textit{i.e.} less toxic) dosing regimen when the increase in gain is negligible. For this reason, we calculate $RG _j$, the relative change of the maximum gain compared to that of each dosing regimen $j$:
\begin{equation} \label{eq:OD-relative_dec}
RG _j = \frac{ G_{max}-G_j}{|G_j|}
\end{equation}
with $G_{max} = \max_j (G_j)$, the gain value associated with the dosing regimen with the maximum gain. This is especially important in our context, where we assume monotonic and increasing relationships for efficacy and PDy activity endpoints, meaning that gain value will decrease only when penalized by safety. This criterion is therefore particularly relevant for supporting plateau detection or small increases of gain.

For the dosing regimen recommendation, different definitions could be considered. One first option is to define the Maximum Gain Dosing regimen within x\% (MGD-x\%) as the lowest dosing regimen for which the relative change in gain (RG) is less than x\%: 
\begin{equation} \label{eq:OD-max_gain_x}
\text{MGD-x\%} = \min \left( \Bigl\{ j \mid RG_j \leq \frac{\text{x}}{100} \Bigl\} \right)
\end{equation}
When $x=0\%$, the MGD-0\% corresponds simply to the lowest dosing regimen with the highest gain value. Then, we can define the Maximum Gain Dosing regimen (denoted as MGD instead of MGD-0\% for ease of notation) as follows: 
\begin{equation} \label{eq:MGD}
\text{MGD} = \min \left( \underset{j}{\mathrm{argmax}}  \,  G_j \right)
\end{equation}
 
 A second option for dosing regimen recommendation is to define the Optimal Dosing regimen within x\% (OD-x\%) as the dosing regimen with the highest probability of being the MGD-x\% . This option allows taking into account the uncertainty surrounding the estimated value of the gain. The probability of being the MGD-x\% for $d_j$ is defined as: 
\begin{equation} \label{eq:proba-opt}
u_j(x) = \Pr\left(j= \min \left( \Bigl\{ k \mid RG_{k} \leq \frac{\text{x}}{100} \Bigl\} \right) \right)
\end{equation}
To obtain the distribution of the gain at each dosing regimen $j$, we used MCMC samples of each model parameter, derived samples of each endpoint $\evalat{p}{D_j}$, $\evalat{q}{D_j}$ and $\evalat{s}{D_j}$, and computed samples of $G_j$. Then, for each sample value, we can identify the MGD-x\%, and estimate the probability of being the MGD-x\% for dosing regimen $j$ as the proportion of MCMC samples where $d_j$ is the MGD-x\%. The OD-x\%, that could be recommended is the one with the highest probability $u_j(x)$ of being the MGD-x\%:
\begin{equation} \label{eq:OD-max_being_opt}
\text{OD-x\%} = \underset{j}{\mathrm{argmax}}  \, u_j(x)
\end{equation}
In this manuscript, we chose $x\% = 1\%$, considering a relative change of gain of 1\% is negligible and does not justify a dose increase.

\subsection{Dose optimization}\label{sec:dose-opt}

As mentioned before, our approach can be employed for cohort dose allocation when safe dosing regimens have been identified (and MTD declared), in order to guide later data collection and information gathering on the different endpoints for the ultimate goal of recommending the phase 2 dosing regimen. In the second part of the dose finding studies, (usually referred to as dose expansion), an adaptive approach can be implemented to allocate patients to optimal dosing candidate. This adaptive approach can be implemented in different ways which may have benefits and hurdles. We foresee two different adaptive approaches, potentially including adaptive randomization, described below.


\begin{algorithm}
\caption{\textit{Multi-step U-DESPE} sequential adaptive design}\label{alg:multistep}
 \begin{enumerate}
    \item Dose-escalation, and MTD determination.
    \item Run U-DESPE to get the probability $u_j(0)$ of each well tolerated dosing regimen being the MGD (cf. Equation (\ref{eq:proba-opt})).
    \item Allocate the next cohort of patients to a dosing regimen selected according to the probability $u_j(0)$ assessed in step (2). For instance, the selected dosing regimen can be the one with the highest probability being the MGD or can be randomly sampled according to their probabilities, $u_j(0)$ of being the MGD, in an adaptive randomization manner.  
    \item Collect data and go back to step (2) until sample size reached.
    \item Final run of U-DESPE and declare the recommended the dosing regimen. 
\end{enumerate}
\end{algorithm}


\textbf{Option 1: \textit{Multi-step U-DESPE} sequential adaptive design}, that would consist of a standard dose-escalation phase followed by an iterative adaptive approach for patient inclusion and dose selection, as described in Algorithm \ref{alg:multistep}. 
This option can be complex to implement operationally and may therefore have some limitations in terms of feasibility.  Indeed, first it requires the PK-model to be updated after each cohort, meaning the PK-samples have to be analyzed and data available after each new cohort of patients, which may be difficult in practice. Other data such as efficacy and PDy activity may also need longer time to be evaluated, which would increase study duration and imply reluctance on such an approach. Feasibility will be further discussed in section \ref{sec:discussion}. 


\begin{algorithm}
\caption{\textit{Two-step U-DESPE} adaptive design with probability-weighted randomization}\label{alg:onestep}
 \begin{enumerate}
    \item Dose-escalation, and MTD determination.
    \item Run U-DESPE to get the probability $u_j(0)$ of each well tolerated dosing regimen being the MGD (cf. Equation (\ref{eq:proba-opt})).
    \item Randomly allocate in parallel the remaining $n$ patients to the well tolerated dosing regimens according to the probability $u_j(0)$ assessed in step (2).
For instance, the number of patients randomized, in parallel, to safe dosing regimen $j=1,...,MTD$ can be 
$\left\lfloor n \times u_j(0) \right\rceil$, where $\left\lfloor . \right\rceil$ is the rounding function to nearest integer. This means that remaining patients are allocated according to probabilities of being the maximum gain dosing regimen, the highest is the probability the more patients are assigned to the dosing regimen. 
    \item Collect data, perform final run of U-DESPE and declare the recommended the dosing regimen. 
 \end{enumerate}
\end{algorithm}

\textbf{Option 2: \textit{Two-step U-DESPE} adaptive design with probability-weighted randomization} that would consist of a standard dose-escalation phase followed by a dose optimization phase, as described in Algorithm \ref{alg:onestep}. This option may be easier to implement in real life and allows for reducing the study duration compared to the \textit{Multi-step U-DESPE} approach. In this option, model updates and estimations from the U-DESPE design will be performed twice: after dose-escalation and at the end of the study.

\section{Simulation study}\label{sec:simu} 
We performed an extensive simulation study to assess the operating characteristics of the proposed method. We considered a first setting where U-DESPE is applied at the end of a dose-escalation phase to recommend the most appropriate dosing regimen, referred to as the \textit{one-step U-DESPE} design. We also considered a second setting where the dose-escalation phase is followed by a dose-optimization phase based on the U-DESPE design to allocate the next patients to optimal dosing regimen candidates. This second setting corresponds to the \textit{two-step U-DESPE} adaptive design with probability-weighted randomization, as described in Algorithm \ref{alg:onestep} in section \ref{sec:dose-opt}. In both cases, a BLRM~\cite{Neuenschwander2008} design was conducted for the dose-escalation phase. \\

Our work was implemented in R version 4.4.1 using Stan \cite{rstan-software:2024}
version 2.32.6 through the “rstan” R package to perform HMC sampling for exposure-endpoints parameters estimation, and using the SAEM (Stochastic approximation of Expectation-Maximization) algorithm through the "nlmixr2" R package to estimate PK parameters. Simulations were performed on the French supercomputer Jean Zay made available to the scientific community by IDRIS (Institute for Development and Resources in Intensive Scientific Computing). For all reported results, we performed 5000 replicates of simulated clinical trials.
 
\begin{figure}[t]
\centerline{\includegraphics[scale=0.75]{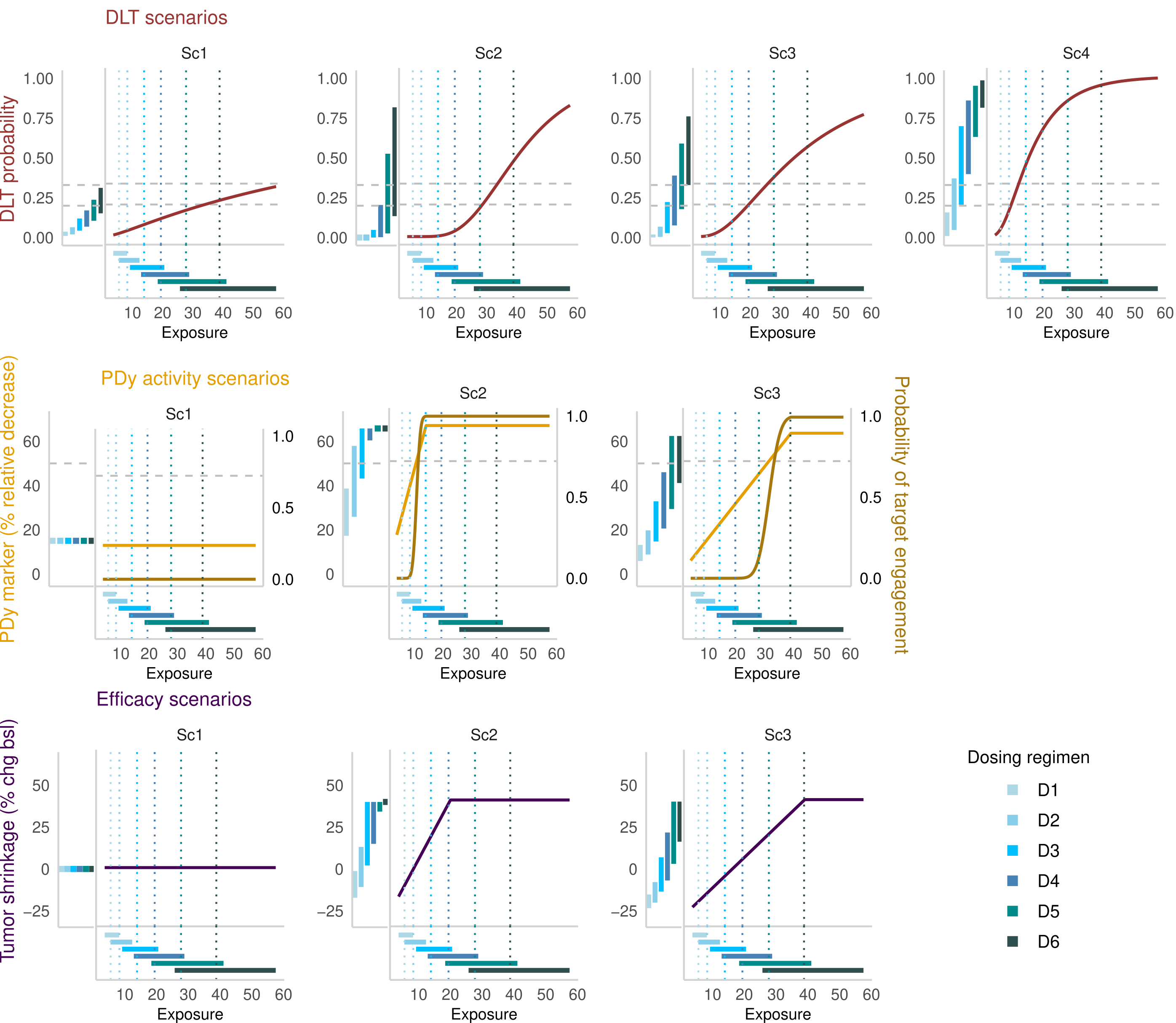}}
\caption{Exposure-endpoint simulated scenarios for safety, efficacy, and PDy activity. For each dosing regimen, the range of exposure is plotted under each scenario (10th and 90th percentiles of the sample exposure) and the possible endpoint value over this range is plotted on the left. Dashed vertical lines represent the median of the sample of the exposure at each dose level.\label{fig:scenarios}}
\end{figure}

 \begin{figure}[!htbp]
\centerline{\includegraphics[scale=0.85]{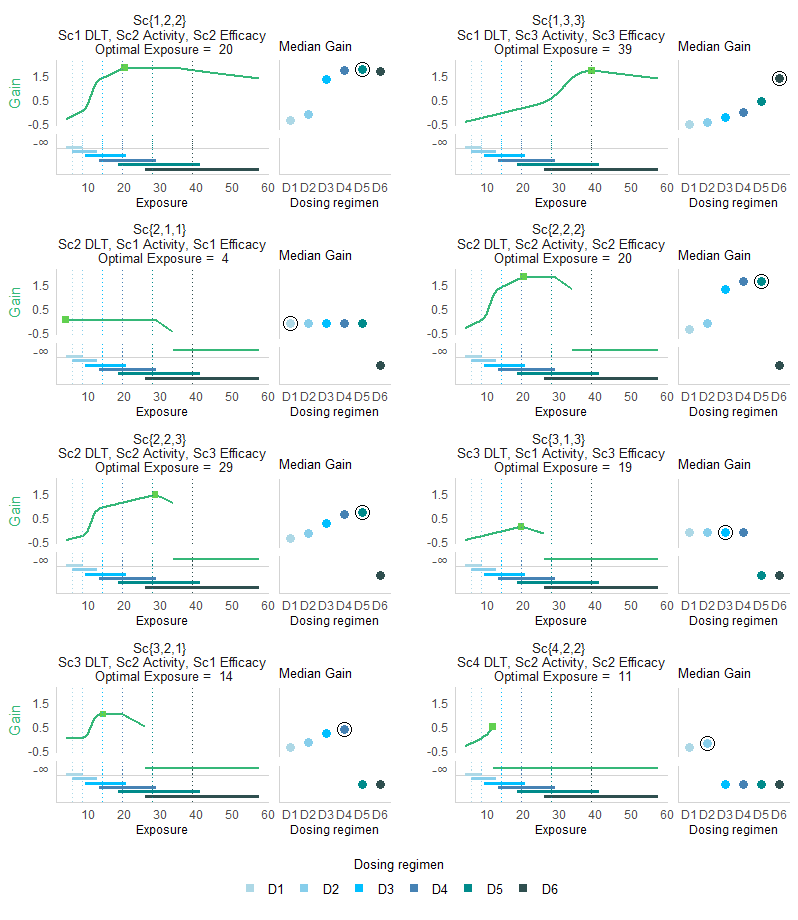}}
\caption{Exposure-Gain relationship for each simulated scenario (in green). Maximum gain is represented by a green square. For each dosing regimen, the range of exposure is plotted under each scenario (10th and 90th percentiles of the sample exposure) and the dashed vertical lines represent the mean of the sample of the exposure at each dosing regimen. Median gain according to dosing regimen is calculated based on the exposure distribution of each dosing regimen, and represented by dots on the right-hand plots.\label{fig:Scenarios_gain_doses}}
\end{figure}

\subsection{Settings}\label{sec:settings}

For all simulations, the total sample size was equal to a maximum of 42 patients. In the first setting, corresponding to the \textit{one-step U-DESPE} design, all 42 patients were allocated during the dose-escalation phase with the BLRM, and the U-DESPE design was applied only at the end. In the second setting, which corresponds to the \textit{two-step U-DESPE} design, in line with our motivating example, 24 patients were allocated during the dose-escalation phase with the BLRM, while the remaining 18 patients were allocated in the dose optimization phase based on the proposed U-DESPE approach. Details on the settings and parameters calibration of the BLRM design can be found in section S1.1 of the Supporting Web Materials.
 
\paragraph{U-DESPE}
The reference exposure, $Z_{ref}$, was 40 for all endpoints. To measure the target engagement, a threshold of $c=0.5$ was specified, meaning that we are interested in pharmacodynamic marker reduction of at least 0.5. For the simulations, utility function parameters were chosen as $(\alpha_1,\alpha_2,\alpha_3,\delta_{\text{min}},\delta_{\text{max}})=(2,1,-4,0.20,0.33)$, where $(\delta_{\text{min}}$ and $\delta_{\text{max}})$ corresponded to the toxicity interval defined for BLRM, and $\alpha_1$ and $\alpha_2$ were the parameters associated with efficacy and PDy activity respectively. In the simulation study, the three endpoints were roughly on the same scale, so the parameter values were chosen according to the importance of each endpoint for dose determination. In practice, these parameters are defined according to the clinician's preferences. Parameters $\alpha_1$ and $\alpha_2$ were set in order to give higher weight on efficacy compared to PDy activity in line with the motivating example, where target engagement is supposed to be necessary for efficacy but not sufficient. To ensure that an increase in the DLT probability after reaching $\delta_{\text{min}}$ will translate into a decrease in utility, we set  $\alpha_3=-4$. In a similar way to that described in section S1.1 of the Supporting Web Materials for BLRM, a weakly informative prior was constructed for DLT, but in terms of exposure rather than dose. Vague priors were defined on the parameters of the PDy activity and efficacy models. We chose a mean around 0 and a large standard error for the Normal distribution on $\gamma_0$ and $\beta$, and vague Gamma priors for $\gamma_l$ and the precisions. To account both for uncertainty around exposure-endpoint models but also around the PK model, we did not assume that PK parameters were known; instead, to stick to reality, we simulated concentrations from which we derived ``true'' exposures to simulate endpoints, and we estimated individual PK parameters to get estimated individual exposures for exposure-endpoints estimations.

\subsection{Simulation scenarios}\label{sec:sim_scenarios}

Our simulation is inspired by the motivating example described in Section~\ref{sec:motivating}. However, due to confidentiality constraints related to an ongoing project, certain details, such as the considered dosing regimens and the PK model, have been modified. We summarized all the details of the data generation process (e.g. parameter value choices) and the simulation steps in Algorithm S1 in the Supporting Web Materials (cf. Section S1.2). We considered $J = 6$ dosing regimens with $L=28$ administrations of repeated doses. In our simulation, each dosing regimen consists of the same dose administered once daily every 24 hours over a 28-day period. PK parameters were chosen to generate exposure variability for each dose level, reflecting the reality of overlap in exposure values between doses.\\

The toxicity, efficacy and PDy activity scenarios are presented in Figure \ref{fig:scenarios}. Different monotone increasing exposure-DLT relationships were considered to explore settings with varying levels of toxicity and different positions for the MTD. Different exposure-PDy and exposure-efficacy relationships were proposed, including null effect and plateau (see section S1.2 of the Supporting Web Materials for details on the construction of the scenarios). These scenarios for each endpoint were combined to build the simulated scenarios according to exposure. The gain function according to exposure for each simulated scenario is provided in Figure \ref{fig:Scenarios_gain_doses}. The first index defining the simulated scenario corresponded to the DLT scenario, the second to the PDy activity scenario, and the third to the efficacy scenario. For each scenario, the 10th and 90th percentiles of the exposure distribution by dose level were plotted on the bottom, and the median of the sample of exposure at each dose level was drawn by dashed vertical lines. In Figure \ref{fig:Scenarios_gain_doses} the true optimal exposure level was provided, defined as the exposure value that maximizes the gain function. Median gain for each dose level was also reported.\\

The simulated scenarios were chosen to encompass different toxicity, PDy activity, and efficacy scenarios resulting in various gain function shapes and optimal exposure positions across the range of candidate dosing regimens. For instance, for scenario \{1,2,2\} all doses were acceptable in terms of toxicity and a plateau occurred from exposure 14 for PDy activity and 20 for efficacy. For low exposure values, the gain function increased as efficacy and PDy activity increased, then remained constant after a plateau was reached for both efficacy and PDy activity at exposure 20, and finally decreased due to toxicity penalty. For this scenario, the optimal exposure was 20, which corresponded to the median exposure of dosing regimen 4, while the dosing regimen associated with the highest median gain across exposure was dosing regimen 5 (dosing regimen 4 being close). Scenarios \{1,2,2\}, \{2,2,2\} and \{3,2,1\} represented targeted therapy scenarios where the correct dose is lower than MTD with different toxicity profiles and gain shapes. Scenarios \{1,3,3\}, \{2,2,3\}, \{3,1,3\} and \{4,2,2\} represented cytotoxic scenarios where the MTD should coincide with the correct dosing regimen with different MTD locations and considering various shapes for the gain function. For scenario \{1,3,3\} the correct dosing regimen is the highest one, while scenario \{4,2,2\} represented a toxic scenario where approximately the second dosing regimen should be correct. Scenario \{3,1,3\} was an example of a scenario where the amplitude of the gain function was small, resulting in more difficult discrimination. Finally, scenario \{2,1,1\} represented the case of no efficacy effect and no PDy activity with dosing regimens; therefore, the lowest dosing regimen should be the correct one far under the MTD.

\subsection{Results}\label{sec:results}
Table \ref{tab:results} presents the simulation results across the eight scenarios. It is important to note that a given dose level can correspond to different exposure values among patients. This variability makes it difficult to recommend a specific dosing regimen with absolute certainty. Due to the overlap in exposure across dosing regimens, different dose levels may yield comparable exposure values, leading to similar efficacy and PDy activity outcomes. For the purposes of this Results section, we define the \textit{correct dosing regimen(s)} for each scenario as the dose with a median exposure closest to the exposure that maximizes the gain function. Additionally, we could consider as a \textit{correct dosing regimen} the dose level that maximizes the median gain calculated based on the exposure distribution of each dosing regimen. These two definitions lead to the same dose recommendation, except for scenarios \{1,2,2\} and \{2,2,2\}. Correct dosing regimens are highlighted in bold in Table \ref{tab:results}.\\

The left-hand side of Table \ref{tab:results} summarizes the simulation results of the \textit{one-step U-DESPE} design for a total sample size of 42 patients, using the BLRM for patient allocation and applying the U-DESPE design to recommend a dose at the end of the trial. Overall, recommendations based on OD-1\% seem more conservative and balanced across the dosing regimens: neighbors of the correct dosing regimen(s) may also be recommended, especially the dosing regimen below. The right-hand side of Table \ref{tab:results} summarizes the simulation results of the \textit{two-step U-DESPE} design corresponding to a dose-escalation phase using a BLRM with 24 patients, followed by a dose-optimization phase of 18 patients using U-DESPE. Under this \textit{two-step U-DESPE} design, with the same total sample size of 42 patients, recommendations based on OD-1\% show significant improvements compared to the \textit{one-step U-DESPE} design: the correct dosing regimen(s) are selected with a much higher percentage, and the tendency to underestimate the correct dosing regimen(s) is effectively eliminated. This can be explained by the fact that the OD metric takes into account the uncertainty surrounding the estimated value of the gain. The dose-optimization phase enables collecting data on candidate correct dosing regimens to discriminate them. On the other hand, recommendations based on MGD-1\% tend to favor a specific dose level and are less spread out across several dose level regimens. In general, correct dosing regimen(s) are identified with a high percentage of correct selection. Moreover, for the same total sample size, the \textit{two-step U-DESPE} design, 
reduced the number of patients treated with dosing regimens higher than the correct dose - and therefore more toxic - compared to the \textit{one-step U-DESPE} design.\\

For scenarios where the correct dose is not necessarily the MTD - because maximum PDy activity and efficacy are achieved at lower dosing regimens - the proposed method reliably recommends the optimal dose. For instance, for the first scenario \{1,2,2\}, where the MTD is at the highest dose level and a plateau in gain is observed approximately at the fourth dosing regimen, the \textit{one-step U-DESPE} design correctly picked the fourth dose in 80.7\% of cases using the MGD-1\% and in 59.1\% of cases using the OD-1\%, while recommending the MTD in only 0.1\% of cases. The fifth dosing regimen can also be considered correct as it maximizes the median gain across exposure distribution, and nearly all patients' exposure reached the gain plateau. This dosing regimen is recommended in 5\% of cases. Under the \textit{two-step U-DESPE} design, 
performance improved. Specifically, when all patients were allocated using BLRM  (\textit{one-step U-DESPE} design), nearly half the patients were allocated to the MTD (i.e. the last dose level), whereas performing a dose-optimization phase based on U-DESPE significantly increased patients' allocation to the optimal doses (dose level 4 and 5), resulting in correct dose recommendations in over 93\% of cases. Additionally, with the \textit{two-step U-DESPE} design, there was less under-estimation of the correct dosing regimen(s), with dosing regimen 5 being recommended more frequently (15.5\% versus 4.4\%). Similar performances are observed for scenario \{2,2,2\} and \{3,2,1\}. When the optimal dose is at the first dose level and MTD at the fourth (scenario \{2,1,1\}), recommendations were more conservative using the OD-1\% compared to the MGD-1\% (99\% versus 33.8\% for the selection of the first dose level). However, regardless of when U-DESPE is applied, dose 1 or 2 is selected in more than 80\% of cases, enabling recommendation of less toxic dose compared to the MTD, while achieving the same efficacy.\\

For scenario \{1,3,3\}, when the correct dose coincides with MTD at the highest dose level, the \textit{one-step U-DESPE} design achieved performance similar to that of BLRM, recommending the correct dose in 75.2\% of cases using the MGD-1\%. Based on the OD-1\% criterion, the proposed method tended to be conservative when applied at the end of the trial (36.5\% of correct selection), while it correctly selected the highest dose when U-DESPE is applied in the middle of the trial to allocate patients (66.3\% of correct selection). As expected, for scenarios with a relatively flat gain function (cf. Figure \ref{fig:Scenarios_gain_doses}), discriminating between doses is more challenging. For instance, for scenario \{3,1,3\}, the \textit{one-step U-DESPE} design hesitated between dosing regimens 3, 4, and 5, with respective selection percentages of 24.5\%, 54.7\% and 16.0\% with OD-1\%.  
However, when U-DESPE was employed during a dose-optimization phase (\textit{two-step U-DESPE} design), the percent of correct selection increased to 61.2\%.  

A contrario, in scenario \{3,2,1\}, with a significant gap between doses in terms of gain, the \textit{one-step U-DESPE} design accurately focused on the correct dose, achieving a percent of correct selection of 90\%. In a toxic scenario where the MTD is at the second dose level (\textit{i.e.} scenario \{4,2,2\}, the U-DESPE design tended to be conservative, by recommending the first dose more often. In 23.2\% of cases, the trial is stopped due to toxicity and no dose is recommended by the BLRM. Trials may also be stopped earlier due to the BLRM accuracy rule, which reduced the number of observed data and made it challenging to estimate models, particularly the PK model parameters. Finally, in the setting of a particularly challenging scenario, such as scenario \{2,2,3\}, with the MTD at dose level 5, high variability in exposure for doses 5 and 6 (cf. scenario 2 of toxicity in Figure \ref{fig:scenarios}), and a plateau in PDy activity and efficacy not reached at the same dosing regimen, the high uncertainty is reflected in the results. In this case, the proposed \textit{two-step U-DESPE} design struggled to discriminate doses, recommending the correct doses in 46.5\% of cases and the two adjacent doses in 29.3\% and 23.9\% of cases. Moreover, in more than 30\% of cases, as BLRM recommended a dose lower than the MTD, then no patients will be assigned to the correct dose during the dose-optimization phase, resulting in complex identification of the correct dose.

\subsection{Sensitivity analysis}\label{sec:sensitivity}

In practice, it can be administratively challenging to allocate patients to many different dosing regimens during the dose optimization phase when dose-escalation is considered over. Therefore, we also performed a sensitivity analysis, where only two dosing regimens were explored during the dose-optimization phase. Patients were allocated according to the two doses with the highest probabilities of being the maximum gain dosing regimens at the end of the escalation phase (according to renormalized probabilities). Table S1
  in the Supporting Material provides the results of this sensitivity analysis. As expected, the operational characteristics are usually preserved, with percentages of correct selection falling within the range of results for patient allocation based solely on the BLRM (i.e. under the \textit{one-step U-DESPE} design) and those observed when both the BLRM and U-DESPE (with dose-optimization performed across all optimal dosing regimen candidates) are used, as shown in Table \ref{tab:results}.

\section{Discussion}\label{sec:discussion}

In this paper, we proposed a unified approach to support dose optimization studies by incorporating safety, pharmacodynamics, efficacy endpoints, and dose-exposure modeling, to guide dosing regimen recommendations. This method can be applied either for final dosing regimen selection at the end of a trial following a conventional DLT-based dose-escalation, or at any stage after exploring safe dose levels, to allocate subsequent cohorts of patients to optimal regimen candidates. The aim of this approach is to identify the optimal dosing regimen based on a comprehensive risk–benefit assessment. As demonstrated in the simulation study across a variety of realistic scenarios, the proposed method appears to be an interesting approach in this setting, exhibiting desirable operating characteristics. When applied at the end of the dose-escalation phase for final dose recommendations, the \textit{one-step U-DESPE} design performed particularly well in selecting the optimal dose, especially in scenarios where the optimal dose was lower than the MTD. This highlights the method's ability to detect plateaus on the gain function and to recommend significantly lower and less toxic doses compared to the MTD, while maintaining comparable PDy activity and efficacy. With the \textit{two-step U-DESPE} design, including a dose-optimization phase after the dose-escalation stage, the percent of correct selection is increased for the same total number of patients. Additionally, fewer patients are allocated to dosing regimens higher than the optimal dose (i.e., more toxic regimens), thereby reducing unnecessary exposure to toxicity. \\

The proposed approach is foreseen to be applied in the framework of a dose finding study or at a later stage, when enough data or information related to the endpoints of interest for dose selection are available (e.g. after a dose-escalation study or for a phase I clinical trial). Indeed, firstly, our approach is based on the assumption that the endpoints may be better explained by the exposure than by the dose, implying the relevant exposure metric should be available for the patients treated as well as for the population.  Population PK modeling is often not relevant before a significant amount of PK data is available. U-DESPE is therefore not planned to be implemented before enough PK data have been analyzed. When they allow the building of a population PK model, the accumulation of new data can be used to refine the distribution of the PK metrics. Secondly, data availability may also be a point of attention for the other endpoints, like endpoints related to pharmacodynamics, or efficacy as these data accumulate with a delay in comparison to safety. For endpoints that may take longer to assess, a potential extension could be to manage partial observations for different endpoints, as suggested by Cheung  \citep{Cheung2000}. However, although a minimum amount of data is required to calibrate the different models, it is not absolutely necessary to have all patients' endpoints evaluable to run the U-DESPE. This is because U-DESPE relies on modeling and relationships can be inferred from the available population data.\\

For simulation purposes, the same PK model was used throughout the trial; however, in practice, this model can be updated based on the most recent information. Moreover, as a global PK model is built for the drug, PK samples collected through different studies can be used to reinforce the knowledge on the pharmacokinetics. Under verifiable assumptions, borrowing relevant information, such as pharmacodynamics data, from different studies via a common analysis can be beneficial and can support a better selection of optimal candidate optimal dosings. Typically, this information sharing between studies is done in a qualitative manner; the U-DESPE methodology allows for a quantitative approach through modeling and analyses. It also facilitates an open and transparent dialogue with health authorities by providing a framework for optimizing dosing regimens. This framework incorporates relevant data and analyses not limited to a single study but extends across the entire program. Furthermore, a major advantage of this approach is that it can compare several dosing regimens with different doses but also with different schedules. Through the exposure with the PK modeling of the compound, the approach may enable choosing the most appropriate combination of dose and schedule.\\

As expected, the distribution of patients to dosing regimens influences the performance of the design in selecting the correct dose. Therefore, different criteria for allocating patients during the dose-optimization phase could be considered. In the simulation study, we allocated the patients according to the probability of being the maximum gain dosing regimen. However, alternative criteria could be explored, for instance, considering the number of cohorts already included at the dosing regimen as in the UCB (Upper Confidence Bounds) algorithm for the bandit model \cite{auer2002, cappe2013}, or allocating patients according to any criterion to maximize information benefit for dose discrimination, thereby increasing the reliability of dose selection. Moreover, U-DESPE is an approach that needs to be tailored based on the relevant endpoints considered for dose optimization. This involves adjusting various pharmacokinetic and exposure-response models, as well as the utility function. Enhancing this approach could involve population exposure-response models. For example, instead of using a basic linear model to describe the best tumor shrinkage per patient, tumor growth inhibition model\cite{Stein2011} could serve as exposure-response models, taking advantage of the information contained in the longitudinal tumor size data\cite{MicallefPoster2017}.\\

Finally, with the molecularly-targeted therapies and immunotherapy agents changing the landscape of dose-finding in oncology, it is essential to use unified approaches for dose recommendation. Exploring multiple dosing regimens during the early clinical development of oncology drugs is essential, with dose optimization based on pharmacodynamy, safety, and efficacy endpoints. In this context, leveraging  modeling and simulation techniques is particularly efficient. By the past, exposure-response analyses were mainly used late in the development to justify or adapt the dose to specific populations (e.g. in pediatrics) while the phase 3 dosage was often selected as the highest tolerated one. With the recent FDA guidance Optimus, the optimization of the dosing regimen needs to occur early to support the dose selection before pivotal phase. The U-DESPE methodology is a tool addressing this requirement, in line with the recommendations from the FDA oncology dosing  toolkit and also offering a robust framework for designing dose-finding trials. Even if this method is meant to be implemented early in the development, it can however be beneficial to any development stage to explore, compare, adapt, select, and justify the use of a given dosing regimen.

\section*{Acknowledgments}
Sandrine Micallef and Anaïs Andrillon made equal contributions and are co‐primary authors. \\
We thank Anne Bellon and Esteban Rodrigo Imedio, Debiopharm International SA, for their contribution during various stages of the paper preparation, comments, and discussions. \\
This work was granted access to the HPC resources of IDRIS under the allocation 20XX-AD010315737 made by GENCI. \\
Part of Moreno Ursino's work was supported by a grant from Inserm and the French Ministry of Health (MESSIDORE 2022, reference number Inserm-MESSIDORE N° 94).

\section*{Conflict of interest}
The authors declare no potential conflict of interests.
\section*{Data availability statement}
Data sharing not applicable to this article as no datasets were generated or analyzed during the current study.

\bibliographystyle{plain}      
\bibliography{biblio}  

@misc{guidance_optimus,
     author = {{US Food and Drug Administration}},
title = {{Optimizing the Dosage of Human Prescription Drugs and Biological Products for the Treatment of Oncologic Diseases: Guidance for Industry}},
  year = {{2024}},
howpublished = {{\url{https://www.fda.gov/regulatory-information/search-fda-guidance-documents/optimizing-dosage-human-prescription-drugs-and-biological-products-treatment-oncologic-diseases}}},
  note = {{Accessed: 2024-08-08}}
}

@article{Thall1998,
author = {Thall, P. F and Russell, K. E},
journal = {Biometrics},
title = {A strategy for dose-finding and safety monitoring based on efficacy and adverse outcomes in phase I/II clinical trials},
volume = {54},
number = {1},
page={251--264},
year = {1998}
}

@misc{project_optimus,
   author = {US Food and Drug Administration},
 title = {Project Optimus: reforming the dose optimization and dose selection paradigm in oncology},
  year = {2021},
  howpublished = {\url{https://www.fda.gov/about-fda/oncology-center-excellence/project-optimus}},
  note = {Accessed: 2023-06-01}
}

@misc{regimen_NCI,
     author = {{National Cancer Insitute}},
title = {NCI Dictionary of Cancer Terms},
  year = {2024},
howpublished = {{\url{https://www.cancer.gov/publications/dictionaries/cancer-terms/def/regimen}}},
  note = {{Accessed: 2024-02-15}}
}

@article{Neuenschwander2008,
author = {Neuenschwander, B and Branson, M and Gsponer, T},
title = {Critical aspects of the Bayesian approach to phase I cancer trials},
journal = {Stat Med},
year = {2008},
volume = {27},
number = {13},
pages = {2420-2439}
}

@article{Debio0123_102,
author = {Papadopoulos, KP and Sharma, M and Dummer, R and Rodrigo Imedio, E and Yge, ML and Micallef, S and Bellon, A and Stathis, A},
title = {A phase 1 dose-finding and dose-expansion study evaluating the safety, tolerability, pharmacokinetics, and efficacy of a highly selective WEE1 inhibitor (Debio 0123) in adult patients with advanced solid tumors},
journal = {J Clin Oncol},
year = {2018},
volume = {40},
number = {16}
}

@misc{Debio0123_101,
 title = {Study of Oral Debio 0123 in Combination With Carboplatin in Participants With Advanced Solid Tumors},
  year = {2019},
  howpublished = {\url{https://www.clinicaltrials.gov/study/NCT03968653}},
  note = {Accessed: 2024-01-26}
}

@article{braun2002,
  title={The bivariate continual reassessment method: extending the CRM to phase I trials of two competing outcomes},
  author={Braun, Thomas M},
  journal={Controlled clinical trials},
  volume={23},
  number={3},
  pages={240--256},
  year={2002},
  publisher={Elsevier}
}

@article{cunanan2014,
  title={Evaluating the performance of copula models in phase I-II clinical trials under model misspecification},
  author={Cunanan, Kristen and Koopmeiners, Joseph S},
  journal={BMC Medical Research Methodology},
  volume={14},
  pages={1--11},
  year={2014},
  publisher={Springer}
}

@article{asakawa2014,
  title={Bayesian model averaging continual reassessment method for bivariate binary efficacy and toxicity outcomes in phase I oncology trials},
  author={Asakawa, Takashi and Hirakawa, Akihiro and Hamada, Chikuma},
  journal={Journal of Biopharmaceutical Statistics},
  volume={24},
  number={2},
  pages={310--325},
  year={2014},
  publisher={Taylor \& Francis}
}

@article{thall2004,
  title={Dose-finding based on efficacy--toxicity trade-offs},
  author={Thall, Peter F and Cook, John D},
  journal={Biometrics},
  volume={60},
  number={3},
  pages={684--693},
  year={2004},
  publisher={Oxford University Press}
}

@article{ivanova2003,
  title={A new dose-finding design for bivariate outcomes},
  author={Ivanova, Anastasia},
  journal={Biometrics},
  volume={59},
  number={4},
  pages={1001--1007},
  year={2003},
  publisher={Oxford University Press}
}

@article{Colin2017,
author = {P. Colin, M. Delattre, P. Minini and S. Micallef},
title = {An Escalation for Bivariate Binary Endpoints Controlling the Risk of Overtoxicity (EBE-CRO): Managing Efficacy and Toxicity in Early Oncology Clinical Trials},
journal = {Journal of Biopharmaceutical Statistics},
volume = {27},
number = {6},
pages = {1054--1072},
year = {2017},
publisher = {Taylor \& Francis}
}

@article{ursino2017,
  title={Dose-finding methods for phase I clinical trials using pharmacokinetics in small populations},
  author={Ursino, Moreno and Zohar, Sarah and Lentz, Frederike and Alberti, Corinne and Friede, Tim and Stallard, Nigel and Comets, Emmanuelle},
  journal={Biometrical Journal},
  volume={59},
  number={4},
  pages={804--825},
  year={2017},
  publisher={Wiley Online Library}
}

@article{micallef2022,
  title={Exposure driven dose escalation design with overdose control: Concept and first real life experience in an oncology phase I trial},
  author={Micallef, Sandrine and Sostelly, Alexandre and Zhu, Jiawen and Baverel, Paul G and Mercier, Francois},
  journal={Contemporary Clinical Trials Communications},
  volume={26},
  pages={100901},
  year={2022},
  publisher={Elsevier}
}

@article{pantoja2022,
author = {Pantoja, Kristyn and Lanke, Shankar and Munafo, Alain and Victor, Anja and Habermehl, Christina and Schueler, Armin and Venkatakrishnan, Karthik and Girard, Pascal and Goteti, Kosalaram},
title = {Designing phase I oncology dose escalation using dose–exposure–toxicity models as a complementary approach to model-based dose–toxicity models},
journal = {CPT: Pharmacometrics \& Systems Pharmacology},
volume = {11},
number = {10},
pages = {1371-1381},
doi = {https://doi.org/10.1002/psp4.12851},
url = {https://ascpt.onlinelibrary.wiley.com/doi/abs/10.1002/psp4.12851},
eprint = {https://ascpt.onlinelibrary.wiley.com/doi/pdf/10.1002/psp4.12851},
year = {2022}
}

@article{takeda2018,
  title={Bayesian dose-finding phase I trial design incorporating pharmacokinetic assessment in the field of oncology},
  author={Takeda, Kentaro and Komatsu, Kanji and Morita, Satoshi},
  journal={Pharmaceutical Statistics},
  volume={17},
  number={6},
  pages={725--733},
  year={2018},
  publisher={Wiley Online Library}
}

@article{gerard2022,
  title={Bayesian dose regimen assessment in early phase oncology incorporating pharmacokinetics and pharmacodynamics},
  author={Gerard, Emma and Zohar, Sarah and Thai, Hoai-Thu and Lorenzato, Christelle and Riviere, Marie-Karelle and Ursino, Moreno},
  journal={Biometrics},
  volume={78},
  number={1},
  pages={300--312},
  year={2022},
  publisher={Oxford University Press}
}

@article{PostelVinay2011 ,
author = {{Postel-Vinay}, S and {Gomez-Roca}, C and Molife, L. R and Anghan, B and Levy, A and Judson, I and {De Bono}, J and Soria, J. C and Kaye, S and Paoletti, X},
journal = {J Clin Oncol},
title = {{Phase I trials of molecularly targeted agents: should we pay more attention to late toxicities?}},
volume = {29},
number = {13},
page={1728--1735},
year = {2011}
}

@article{Stein2011,
author = {Stein, WD and Gulley, JL and Schlom, J and Madan, RA and Dahut, W and Figg, WD and Ning, YM and Arlen, PM and Price, D and Bates, SE and Fojo, T},
journal = {Clin Cancer Res},
title = {{Tumor regression and growth rates determined in five intramural NCI prostate cancer trials: the growth rate constant as an indicator of therapeutic efficacy}},
volume = {17},
number = {4},
page={907--17},
year = {2011}
}

@inproceedings{MicallefPoster2017,
author = {Micallef, S and Mercier, F},
title= {Evaluation of tumor kinetics metrics as early endpoint to support decision making in early drug development},
booktitle   = {},
year= {2017},
publisher = {Poster at PAGES},
pages     = {},
url       = {https://www.page-meeting.org/pdf_assets/3658-PAGE%202017%20micallef%20final.pdf}
}

@Misc{rstan-software:2024,
    title = {{RStan}: the {R} interface to {Stan}},
    author = {{Stan Development Team}},
    note = {R package version 2.32.6},
    year = {2024},
    url = {https://mc-stan.org/},
}

@article{su2022,
  title={A semi-mechanistic dose-finding design in oncology using pharmacokinetic/pharmacodynamic modeling},
  author={Su, Xiao and Li, Yisheng and M{\"u}ller, Peter and Hsu, Chia-Wei and Pan, Haitao and Do, Kim-Anh},
  journal={Pharmaceutical statistics},
  volume={21},
  number={6},
  pages={1149--1166},
  year={2022},
  publisher={Wiley Online Library}
}

@article{yang2024,
  title={An extended Bayesian semi-mechanistic dose-finding design for phase I oncology trials using pharmacokinetic and pharmacodynamic information},
  author={Yang, Chao and Li, Yisheng},
  journal={Statistics in Medicine},
  volume={43},
  number={4},
  pages={689--705},
  year={2024},
  publisher={Wiley Online Library}
}

@article{yuan2024,
  title={Pharmacometrics-Enabled DOse OPtimization (PEDOOP) for seamless phase I-II trials in oncology},
  author={Yuan, Shijie and Huang, Zhanbo and Liu, Jiaxin and Ji, Yuan},
  journal={Journal of Biopharmaceutical Statistics},
  pages={1--20},
  year={2024},
  publisher={Taylor \& Francis}
}

@article{yuanOpt2024,
  title={Statistical and practical considerations in planning and conduct of dose-optimization trials},
  author={Yuan, Ying and Zhou, Heng and Liu, Suyu},
  journal={Clinical trials },
  volume={21},
  number={3},
  pages={273--286},
  year={2024},
  publisher={Elsevier}
}

@article{Lu2024,
  title={A Bayesian pharmacokinetics integrated phase I-II design to optimize dose-schedule regimes},
  author={Lu, M and Yuan, Y and Liu, S},
  journal={Biostatistics},
  volume={26},
  number={1},
  pages={kxae034},
  year={2024} 
}

@article{Hoering2011,
  title={Seamless phase I-II trial design for assessing toxicity and efficacy for targeted agents},
  author={ Hoering, Antje and LeBlanc, Michael and Crowley, John},
  journal={ Clinical cancer research},
  volume={17},
  number={4},
  pages={640--646},
  year={2011},
  publisher={Taylor \& Francis}
}

@article{Guo2023,
  title={DROID: dose-ranging approach to optimizing dose in oncology drug development. Biometrics},
  author={Guo, Beibei and Yuan, Ying},
  journal={Biometrics},
  volume={79},
  number={4},
  pages={2907--2919},
  year={2023},
  publisher={Wiley Online Library}
}

@article{Cheung2000,
  title={Sequential designs for phase I clinical trials with late-onset toxicities},
  author={Cheung, Ying Kuen and Chappell, Rick},
  journal={Biometrics},
  volume={56},
  number={4},
  pages={1177--1182},
  year={2000},
  publisher={Jstor}
}

@article{auer2002,
  title={Finite-time Analysis of the Multiarmed Bandit Problem},
  author={Auer, P and Cesa-Bianchi, N. and Fischer, P},
  journal={Machine Learning},
  volume={47},
  number={2},
  pages={235–256},
  year={2002}
}

@article{cappe2013,
  title={Kullback-Leibler upper confidence bounds for optimal sequential allocation},
  author={Capp{\'e}, Olivier and Garivier, Aur{\'e}lien and Maillard, Odalric-Ambrym and Munos, R{\'e}mi and Stoltz, Gilles},
  journal={The Annals of Statistics},
  pages={1516--1541},
  year={2013},
  publisher={JSTOR}
}

@article{ramsay1988,
  title={Monotone regression splines in action},
  author={Ramsay, James O},
  journal={Statistical science},
  pages={425--441},
  year={1988},
  publisher={JSTOR}
}

 \begin{landscape}
\begin{table}
\begin{center}
\caption{Simulation results of, on left, \textit{one-step U-DESPE} design (U-DESPE applied after a BLRM with 42 patients), and on right, \textit{two-step U-DESPE} design (BLRM with 24 patients followed by a dose-optimization phase based on U-DESPE of 18 additional patients). Results are given terms of percent of selection at each dosing regimen and average number of patients. Approximate correct selection results based on optimal dosing regimen are given in boldface and those of MTD are underlined.\label{tab:results}}
\begin{tabular}{lllccccccccccp{0.5cm}ccccccc} 
\toprule
 &  &  & \multicolumn{7}{c}{\textit{One-step U-DESPE design}} & & \multicolumn{7}{c}{\textit{Two-step U-DESPE design}} \\ \cline{4-10} \cline{12-18} 
 \multicolumn{2}{l}{\textbf{Dosing regimens}} &  & \textbf{Stop} & \multicolumn{1}{c}{\textbf{1}} & \textbf{2} & \textbf{3} & \textbf{4} & \textbf{5} & \textbf{6} & & \textbf{Stop} & \textbf{1} & \textbf{2} & \textbf{3} & \textbf{4} & \textbf{5} & \textbf{6} \\ \hline

Sc\{1,2,2\}  & BLRM & \% MTD& 0.2 & 0.0 & 0.0 & 0.2 & \textbf{3.9} & \textbf{19.8} & \underline{75.9} && 0.2 & 0.0 & 0.1 & 2.0 & \textbf{10.1} & \textbf{24.2} & \underline{63.4} \\
 & U-DESPE & \% MGD-1\% &  & 0.0 & 0.0 & 14.6 & \textbf{80.7} & \textbf{4.4} & 0.1 &&  & 0.0 & 0.0 & 0.9 & \textbf{82.7} & \textbf{15.5} & 0.6 \\
 &  & \% OD-1\% &  & 0.0 & 0.0 & 35.1 & \textbf{59.1} & \textbf{5.4} & 0.1 &&  & 0.0 & 0.0 & 5.1 & \textbf{77.9} & \textbf{15.7} & 1.0 \\ 
 \cline{3-18} 
 &  & No patients &  & 3.0 & 3.3 & 3.6 & 5.4 & 8.1 & 18.3 &&  & 3.1 & 3.6 & 5.0 & 8.2 & 10.9 & 11.0 \\
\hline

Sc\{1,3,3\} & BLRM & \% MTD & 0.2 & 0.0 & 0.0 & 0.2 & 3.9 & 19.8 & \textbf{\underline{75.9}} && 0.2 & 0.0 & 0.1 & 2.0 & 10.1 & 24.2 & \textbf{\underline{63.4}} \\
 & U-DESPE & \% MGD-1\% &  & 0.0 & 0.1 & 1.2 & 3.1 & 20.3 & \textbf{75.2} &&  & 0.2 & 0.1 & 0.9 & 5.4 & 20.5 & \textbf{72.6} \\
 &  & \% OD-1\% &  & 0.0 & 0.0 & 0.8 & 7.2 & 55.2 & \textbf{36.5} &&  & 0.2 & 0.0 & 0.2 & 5.9 & 27.2 & \textbf{66.3} \\ 
 \cline{3-18}
  &  & No patients &  & 3.0 & 3.3 & 3.6 & 5.4 & 8.1 & 18.3 && & 3.3 & 3.8 & 4.5 & 7.0 & 9.8 & 13.6 \\
\hline

Sc\{2,1,1\} & BLRM & \% MTD & 0.0 & \textbf{0.0} & 0.0 & 0.1 & 19.4 & \underline{66.7} & 13.8 && 0.0 & \textbf{0.0} & 0.0 & 1.6 & 29.1 & \underline{46.2} & 23.1 \\
& U-DESPE & \% MGD-1\% &  & \textbf{33.8} & 61.5 & 4.3 & 0.3 & 0.1 & 0.0 &&  & \textbf{7.9} & 75.4 & 15.0 & 1.5 & 0.2 & 0.0 \\
 &  & \% OD-1\% &  & \textbf{99.9} & 0.0 & 0.0 & 0.0 & 0.0 & 0.0 &&  & \textbf{99.9} & 0.0 & 0.0 & 0.0 & 0.0 & 0.1 \\ 
 \cline{3-18} 
 &  & No patients &  & 3.0 & 3.0 & 3.4 & 9.0 & 14.9 & 8.0 &&  & 7.4 & 4.8 & 7.0 & 8.2 & 8.7 & 5.8 \\
\hline

Sc\{2,2,2\} & BLRM & \% MTD & 0.0 & 0.0 & 0.0 & 0.1 & \textbf{19.4} & \textbf{\underline{66.7}} & 13.8 && 0.0 & 0.0 & 0.0 & 1.6 & \textbf{29.1} & \textbf{\underline{46.2}} & 23.1 \\
& U-DESPE & \% MGD-1\% &  & 0.0 & 0.0 & 20.0 & \textbf{79.2} & \textbf{0.8} & 0.0 &&  & 0.0 & 0.0 & 1.2 & \textbf{90.2} & \textbf{8.4} & 0.1 \\
 &  & \% OD-1\% &  & 0.0 & 0.0 & 24.1 & \textbf{73.8} & \textbf{2.2} & 0.0 &&  & 0.0 & 0.0 & 1.6 & \textbf{78.6} & \textbf{19.5} & 0.2 \\ 
 \cline{3-18} 
 &  & No patients &  & 3.0 & 3.0 & 3.4 & 9.0 & 14.9 & 8.0 &&  & 3.1 & 3.6 & 6.5 & 12.1 & 10.8 & 5.9 \\
\hline

Sc\{2,2,3\} & BLRM & \% MTD & 0.0 & 0.0 & 0.0 & 0.1 & 19.4 & \textbf{\underline{66.7}} & 13.8 && 0.0 & 0.0 & 0.0 & 1.6 & 29.1 & \textbf{\underline{46.2}} & 23.1 \\
& U-DESPE & \% MGD-1\% &  & 0.0 & 0.0 & 3.6 & 21.7 & \textbf{72.2} & 2.6 &&  & 0.0 & 0.0 & 0.4 & 29.3 & \textbf{46.5} & 23.9 \\
 &  & \% OD-1\% &  & 0.0 & 0.0 & 1.9 & 39.9 & \textbf{53.4} & 4.8 &&  & 0.0 & 0.0 & 0.0 & 26.9 & \textbf{37.2} & 35.8 \\ 
 \cline{3-18} 
 &  & No patients &  & 3.0 & 3.0 & 3.4 & 9.0 & 14.9 & 8.0 &&  & 3.3 & 3.8 & 5.7 & 10.5 & 11.6 & 6.9 \\
\hline

Sc\{3,1,3\} & BLRM & \% MTD & 0.0 & 0.0 & 0.0 & 9.5 & \textbf{\underline{58.6}} & 30.3 & 1.6 && 0.0 & 0.0 & 1.3 & 16.8 & \textbf{\underline{46.6}} & 28.6 & 6.7 \\
& U-DESPE  & \% MGD-1\% &  & 1.8 & 2.4 & 38.2 & \textbf{29.7} & 27.3 & 0.6 &&  & 1.9 & 0.8 & 26.5 & \textbf{46.3} & 21.1 & 3.5 \\
 &  & \% OD-1\% &  & 2.6 & 1.5 & 24.5 & \textbf{54.7} & 16.0 & 0.7 &&  & 1.8 & 0.0 & 11.2 & \textbf{61.2} & 21.5 & 4.1 \\ 
 \cline{3-18} 
 &  & No patients &  & 3.0 & 3.7 & 7.0 & 15.0 & 9.6 & 2.5 &&  & 4.4 & 6.1 & 8.8 & 12.3 & 7.9 & 2.5 \\
\hline

Sc\{3,2,1\} & BLRM & \% MTD & 0.0 & 0.0 & 0.0 & \textbf{9.5} & \underline{58.6} & 30.3 & 1.6 && 0.0 & 0.0 & 1.3 & \textbf{16.8} & \underline{46.6} & 28.6 & 6.7 \\
& U-DESPE & \% MGD-1\% &  & 0.1 & 0.3 & \textbf{92.2} & 7.4 & 0.0 & 0.0 &&  & 0.0 & 0.0 & \textbf{76.8} & 23.1 & 0.0 & 0.0 \\
 &  & \% OD-1\% &  & 1.9 & 7.6 & \textbf{90.5} & 0.0 & 0.0 & 0.0 &&  & 0.1 & 0.0 & \textbf{99.8} & 0.0 & 0.0 & 0.0 \\ 
 \cline{3-18} 
 &  & No patients &  & 3.0 & 3.7 & 7.0 & 15.0 & 9.6 & 2.5 &&  & 4.9 & 6.9 & 11.2 & 10.7 & 6.2 & 2.1 \\
\hline

Sc\{4,2,2\} & BLRM & \% MTD & 23.2 & 18.2 & \textbf{\underline{55.8}} & 2.7 & 0.0 & 0.0 & 0.0 && 25.5 & 20.1 & \textbf{\underline{48.3}} & 5.8 & 0.3 & 0.0 & 0.0 \\
& U-DESPE  & \% MGD-1\% &  & 57.0 & \textbf{13.5} & 2.7 & 0.0 & 0.0 & 0.0 &&  & 65.2 & \textbf{9.3} & 0.1 & 0.0 & 0.0 & 0.0 \\
 &  & \% OD-1\% &  & 57.6 & \textbf{14.5} & 1.1 & 0.0 & 0.0 & 0.0 &&  & 64.7 & \textbf{9.7} & 0.1 & 0.0 & 0.0 & 0.0 \\ 
 \cline{3-18} 
 &  & No patients &  & 10.0 & 17.1 & 3.6 & 0.8 & 0.1 & 0.0 &&  & 15.7 & 13.7 & 3.4 & 0.8 & 0.1 & 0.0 \\
\bottomrule%
\end{tabular}
\end{center}
\end{table}
\end{landscape}

\section*{Supporting information}\label{sec:sup_inf}
\renewcommand\thefigure{S\arabic{figure}} 
\renewcommand\thetable{S\arabic{table}} 
\renewcommand\thesection{S\arabic{section}} 
\renewcommand\thealgorithm{S\arabic{algorithm}} 
\setcounter{section}{0}       
\setcounter{figure}{0}        
\setcounter{table}{0}         
\setcounter{algorithm}{0}     

\section{Simulation study}
\subsection{Settings}
\paragraph{BLRM}
No dosing regimen skipping was allowed during escalation; the first dosing regimen was considered as the starting dose, and patients were included by cohorts of 3. The reference dose was chosen as 50 mg. To guide dose-escalation, inspired by our motivating example, we defined the target toxicity interval $[\delta_{min}; \delta_{max}]=[0.2;0.33]$. The overdose control rule considered a dose as unsafe if the observed data suggested that there is a 25\% or higher posterior probability that the DLT rate of a dosing regimen is greater than $\delta_{max}$. 
A weakly informative prior, as proposed by Neuenschwander \cite{Neuenschwander2008}, was used. The risk of underdosing at the lowest dose was set to 90\% (\textit{i.e.} $\Pr(p_1<\delta_{min})=0.9$) and the risk of an acceptable toxicity at the highest dose was set up to 20\% (\textit{i.e.} $\Pr(p_J<\delta_{max})=0.20$). The dose-escalation was terminated either for accuracy, or when the dose-escalation sample size was reached, or if all dosing regimens were considered too toxic. Accuracy was settled if the recommended dosing regimen was already administered to the last 3 consecutive cohorts of patients and if $\Pr(\delta_{min}<p^{*}<\delta_{max})\geq 0.60$). For the case when all dosing regimens were estimated overtoxic, the U-DESPE was not applied whatever the setting and no dosing regimen was recommended.\\

\subsection{Data generation and scenarios}\label{sec:data_generation}

We considered $J = 6$ dosing regimens with $L=28$ administrations of repeated doses: $D_j = \left( d_j^{(1)}, ..., d_j^{(L)} \right)$ for $j=1,...,J$. In our simulations, each dosing regimen consists of the same dose administered once daily every 24 hours over a 28-day period, that is $\forall \ell \in [1,L]$, $d_j^{(\ell)}=d_j$, and $D_j = [(d_j)_{\times L}]$ with $(d_1, d_2, d_3, d_4, d_5, d_6) = (10, 15, 25, 35, 50, 70)$ mg.

\paragraph{PK generation}
 The PK model presented in equation (2) in the main paper was consistent across all simulation scenarios with the following parameter values: $k_{a,pop} = 1$, $CL_{pop} = 1.8$, $V_{pop} = 100$, $\sigma_{k_a}^2 = 0.3$, and $\sigma_{CL}^2 = 0.1$. PK parameters were chosen to generate exposure variability for each dose level, reflecting the reality of overlap in exposure values between doses. The corresponding distribution of exposure for each dosing regimen is plotted in Figure \ref{fig:Dose_exposure}. The AUC following each drug administration was computed using the actual $i$-th patient’s blood concentration, $f(\theta_i, t, D_i)$. However, for the U-DESPE method, noisy simulated drug concentrations, $c_{ik}$, in blood plasma were used:  
\[
c_{ik} = f(\theta_i, d_i, t_{k}) (1 + \epsilon_{ik}),
\]
where $\epsilon_{ik} \sim N(0, 0.1^2)$ is the error term, and measurements were taken at time $t_{k}$. To mimic the motivating example, each simulated patient could have a maximum of 19 blood concentration samples at predefined time points: $t_k \in \{0.5, 1, 2, 3, 4, 6, 8, 23.5, 25, 26, 27, 28, 30, 32, 47, 169, 176, 337, 344\}$ hours post-first drug intake. If a DLT occurred, the drug schedule was interrupted, and subsequent blood sampling was also discontinued.

\begin{figure}[t]
\centerline{\includegraphics[scale=0.75]{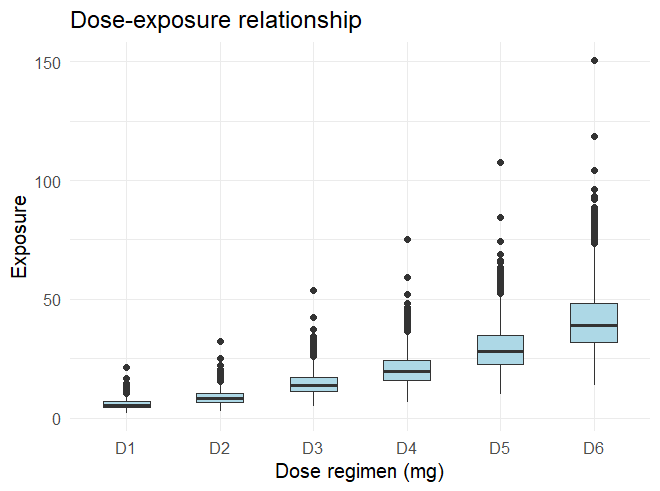}}
\caption{Distribution of exposure for each dosing regimen assessed using the population PK model. The lower and upper hinges correspond to the first and third quartiles (the 25th and 75th percentiles).}\label{fig:Dose_exposure}
\end{figure}

\paragraph{\textbf{DLT simulation}}

To simulate DLT from exposure profile, we defined a threshold $\tau$ on the exposure and assumed that DLT occurred if this threshold was exceeded (\cite{ursino2017, gerard2022}). To introduce between subject variability, we defined a log-normally distributed measure of subject sensitivity, $\kappa_i$, for patient $i$, where $\kappa_i = e^{\eta_{\kappa_i}}$ and $\eta_{\kappa_i} \sim \mathcal{N}(0,\omega^2_{\kappa})$. We assumed that patient $i$ experienced DLT at the $\ell$th administration ($W_{i,\ell}=1$) if $\kappa_i Z_{i,\ell} \geq \tau$ where $Z_{i,\ell}$ is the exposure of patient $i$ at the $\ell$th administration. To mimic a real clinical trial, we assumed that once a DLT is observed, administrations are stopped. If no DLT was observed until the end of the administration cycle that is $\forall \ell, W_{i,\ell}=0$, then the patient did not experience a DLT, $W_i=0$, otherwise $W_i=1$. The true exposure metric considered for our simulations and derived from generated true concentrations, AUC24 (area under the curve over the last 24 hours starting after last administration), is then updated considering that administrations are stopped when a DLT is observed.\\ 

Different monotone increasing exposure-DLT relationships are considered to explore settings with varying levels of toxicity and different positions for the MTD, as shown in Figure 3 in the main paper. The first toxicity scenario was defined to ensure that all doses are safe with $(\omega_{\kappa}=1.5, \tau=120)$, contrary to the fourth scenario where almost all doses are toxic with $(\omega_{\kappa}=0.9, \tau=14)$. Two intermediate scenarios were also defined: scenario 2 with $(\omega_{\kappa}=0.7, \tau=35)$ and scenario 3 with $(\omega_{\kappa}=0.9, \tau=14)$.

\paragraph{\textbf{PDy activity and Efficacy generation}}
Efficacy and PDy activity outcomes were simulated from a normal distribution, with the mean defined by a linear relationship with the exposure $Z_i$ (\textit{i.e.}, $\upsilon_0 + \upsilon_1 Z_i$) and a standard deviation of 0.05 to account for patient variability. To incorporate a plateau effect beyond a specified exposure value $Z_0$, the mean of the normal distribution was fixed at $\upsilon_0 + \upsilon_1 Z_0$ for all exposure values exceeding $Z_0$.\\

The PDy activity and efficacy scenarios are illustrated in Figure 3 in the main paper. For efficacy, scenario 1 represents a null effect with no tumor shrinkage $(\upsilon_0 = Z_0 = 0)$. Scenarios 2 and 3 model plateau exposure-efficacy relationships with a plateau occurring at different dosing regimens, and parameters set to $(\upsilon_0 = -0.3, \upsilon_1 = 0.035, Z_0 = 20)$ and $(\upsilon_0 = -0.3, \upsilon_1 = 0.018, Z_0 = 39)$, respectively. PDy activity scenarios were defined analogously: $(\upsilon_0 = 0.15, \upsilon_1 = 0.045, Z_0 = 0)$ for scenario 1, $(\upsilon_0 = 0, \upsilon_1 = 0.047, Z_0 = 14)$ for scenario 2, and $(\upsilon_0 = 0, \upsilon_1 = 0.016, Z_0 = 39)$ for scenario 3.\\

We summarized all the simulation steps in Algorithm \ref{alg:simu}.

\begin{algorithm}
\caption{Simulations algorithm}\label{alg:simu}
\begin{algorithmic}[1]
\Statex For a large number of clinical trials:
\begin{enumerate}
    \item Simulate a dose-escalation phase with a BLRM design and determine the MTD. For each cohort of patients:
    \begin{itemize}
        \item Simulate concentrations and derive true exposure metrics at different time points from current dosing regimen 
        \item Simulate time of DLT occurrence and DLT from true exposure metric accounting for patient's variability, and derived the true observed exposure metrics that is exposure metrics until a DLT is observed or through the end of the administration cycle
        \item Simulate efficacy and pharmacodynamic data from true observed exposure metrics accounting for patient's variability
        \item Determine the next recommended dose or the MTD from DLT and administered dosing regimens data only
    \end{itemize}
    \item  Estimate PK model with concentrations (until a DLT is observed or through the end of the administration cycle) and administered dosing regimen data as explained in 3.1 in the main paper
    \begin{itemize}
        \item Estimate observed individual exposure metrics for each patient (until a DLT is observed or through the end of the administration cycle)
        \item Provide large samples of exposure metrics per dosing regimen from estimated population PK model
    \end{itemize}
    \item Estimate exposure-endpoint models with estimated exposure metrics, DLT, efficacy and PDy activity data as explained in 3.2, 3.3, and 3.4 in the main paper
 \item Derive dosing regimen-endpoint estimations with estimated population PK samples of exposure metrics per dosing regimen as explained 3.5 in the main paper
    \item Estimate utility at each dosing regimen as explained in 3.6 in the main paper
    \item Estimate probabilities of being the MGD-x\%, and determine the MGD-x\% and the OD-x\% as explained in 3.7 in the main paper
    \item[(*)] If $n$ additional patients can be included after the end of BLRM according to U-DESPE design:  
    \begin{itemize}
        \item Allocate in parallel $\left\lfloor n \times u_j(0) \right\rceil$ patients to dosing regimens $j=1,...,MTD$  as explained in 3.8 in the main paper
        \item Simulate concentrations, DLT, efficacy and PDy activity data at dosing regimens as in (1)
        \item Repeat estimations from (2) to (6) to determine the MDG-x\% and OD-x\%
    \end{itemize} 
\end{enumerate}
\end{algorithmic}

\end{algorithm}

\newpage
\section{Sensitivity analysis}\label{sec:Sensitivity}

  \begin{center}
 
\begin{longtable}{lllccccccc} 
\caption{Global simulation results with 5000 replicated trials of U-DESPE design applied after a dose-escalation phase of a BLRM with 24 patients and a dose-optimization phase of 18 additional patients according to U-DESPE for a total of 42 patients. For the dose-optimization phase, only two dosing regimens were explored, corresponding to the doses associated with the two best probabilities of being the maximum gain dosing regimen at the end of the escalation phase. Approximate correct selection results based on optimal dosing regimen are given in boldface those of MTD are underlined.\label{tab:results_appendix}}\\
\toprule
 &  &  &  & \multicolumn{6}{c}{Dosing regimens}   \\ \cline{5-10} 
 Sc &  &  & \textbf{Stop} & \multicolumn{1}{c}{\textbf{1}} & \textbf{2} & \textbf{3} & \textbf{4} & \textbf{5} & \textbf{6} \\ \hline
\{1,2,2\}  & BLRM & \% MTD&  0.2 & 0.0 & 0.1 & 2.0 & \textbf{10.1} & \textbf{24.2} & \underline{63.4}\\
  & U-DESPE & \% MGD-1\%& & 0.200 & 0.0 & 7.5 &\textbf{83.2} & \textbf{8.8} & 0.4 \\
 &  & \% OD-1\% &&  0.0  &0.0& 20.9 &\textbf{66.9}& \textbf{11.1} & 0.8\\ 
 \cline{3-10}
 &  & No patients &&   3.1 & 3.5 & 4.4 & 7.3 & 11.9 &11.7\\

\hline

\{1,3,3\} & BLRM & \% MTD & 0.2 & 0.0 & 0.1 & 2.0 &10.1 &24.2& \textbf{\underline{63.4}} \\
 
 & U-DESPE & \% MGD-1\% &  &0.2   &0.1  & 1.2  & 4.2  &18.7  &\textbf{75.3} \\
 &  & \% OD-1\% &  &0.2  &0.0 &  0.1  & 5.4 & 37.8   &\textbf{56.3} \\ 
 \cline{3-10}
  &  & No patients  &  & 3.1 &  3.5 &  4.3  & 6.7 & 10.2  &14.1\\
\hline

\{2,1,1\} & BLRM & \% MTD &  0.0 &  \textbf{0.0}&  0.0 & 1.6 &29.1& \underline{46.2} &23.1\\
 
 &  & \% MGD-1\% && \textbf{31.7}& 62.9 & 5.2 & 0.2 & 0.0 & 01 \\
 &  & \% OD-1\% & &  \textbf{100}  &0.0 & 0.0 &0.0 & 0.0 & 0.0 \\ 
 \cline{3-10}
 &  & No patients & &  9.1& 3.1&  5.1 & 6.9& 11.1 & 6.7 \\
\hline

\{2,2,2\} & BLRM & \% MTD &0.0 &  0.0 & 0.0  &1.6& \textbf{29.1}&\textbf{\underline{46.2}}& 23.1\\
 & U-DESPE  & \% MGD-1\% &&  0.0 & 0.0 & 9.5 &\textbf{81.4} & \textbf{9.2} & 0.0\\
 &  & \% OD-1\% & &  0.0 & 0.0 &12.9 &\textbf{73.9}& \textbf{13.1 }& 0.1 \\ 
 \cline{3-10}
 &  & No patients &&  3.0 & 3.2 & 5.7& 12.2& 11.6 & 6.2  \\

\hline

\{2,3,2\} & BLRM & \% MTD & 0.0 &  0.0 & 0.0 & 1.6&29.1& \textbf{\underline{46.2}} &23.1\\
 
 &  & \% MGD-1\% & & 0.0  &0.0&  0.4 &27.2& \textbf{63.5} & 8.8\\
 &  & \% OD-1\% &  &  0.0 & 0.0 & 0.20 &24.2 &\textbf{61.7}& 13.9\\ 
 \cline{3-10}
 &  & No patients &  &  3.0 & 3.3 & 5.5& 10.9& 12.2&  7.1  \\

\hline

\{3,3,1\} & BLRM & \% MTD & 0.0 & 0.0 & 1.3 &16.8& \textbf{\underline{46.6}} &28.6&  6.68 \\
 & U-DESPE  & \% MGD-1\% && 1.6  &0.6& 39.6& \textbf{42.8 }&14.8 & 0.5  \\
 &  & \% OD-1\% & &1.6 & 0.1 &17.1 &\textbf{64.6} &16.1 & 0.5 \\ 
 \cline{3-10}
 &  & No patients & & 3.3  &5.5&  9.2 &13.2 & 8.2 & 2.5\\

\hline

\{3,1,2\} & BLRM & \% MTD & 0.0 & 0.0 & 1.3 &\textbf{16.8} &\underline{46.6} & 28.6 & 6.7 \\
 
& U-DESPE  & \% MGD-1\% & & 0.0 & 0.1 &\textbf{87.9}& 11.9 & 0.0 & 0.0\\
 &  & \% OD-1\% &  & 0.7 & 2.2& \textbf{97.1} & 0.0 & 0.0 & 0.0\\ 
 \cline{3-10}
 &  & No patients & & 3.3 & 5.4 &12.3& 11.5  &7.2 & 2.4\\

\hline

\{4,2,2\} & BLRM & \% MTD &25.5 & 20.1 &\textbf{\underline{48.3}} & 5.8 & 0.3 & 0.0 & 0.0  \\
 
& U-DESPE  & \% MGD-1\% & &65.7  &\textbf{8.8 } &0.1 & 0.0  &0.0 & 0.0 \\
 &  & \% OD-1\% & & 65.3  &\textbf{9.2 } &0.1 & 0.0 & 0.0 & 0.0 \\ 
 \cline{3-10}
 &  & No patients & & 15.6 &13.8  &3.5  &0.8 & 0.1 & 0.0 \\

\bottomrule
\end{longtable}
\end{center}


\end{document}